\documentclass[aps,prl,10pt,twocolumn,amsmath,amssymb,floatfix,superscriptaddress,longbibliography]{revtex4-2}

\usepackage{graphicx} % include figure files
\usepackage{bm} % bold math

\usepackage{hyperref}

\usepackage{braket,bbold,color,enumerate}

\newcommand{\Jdag}{J^{\dag}}

\newcommand{\cdag}{c^{\dag}}

\newcommand{\Hpost}{H_{\mathrm{post}}}

\newcommand{\kket}[1]{|#1\rangle\!\rangle}

\usepackage{geometry}
\usepackage{amsmath,amssymb,xcolor}
\usepackage{amsfonts}
\usepackage{amsmath}
\usepackage{mathtools}
\usepackage{dsfont}

\usepackage{xcolor}

\begin{document}
\title{Lindbladian versus Postselected Non-Hermitian Topology}

\author{Alexandre Chaduteau}
\affiliation{Blackett Laboratory, Imperial College London, London SW7 2AZ, United Kingdom}

\author{Derek K. K. Lee}
\affiliation{Blackett Laboratory, Imperial College London, London SW7 2AZ, United Kingdom}

\author{Frank Schindler}
\affiliation{Blackett Laboratory, Imperial College London, London SW7 2AZ, United Kingdom}

\begin{abstract}
The recent topological classification of non-Hermitian `Hamiltonians' is usually interpreted in terms of pure quantum states that decay or grow with time. However, many-body systems with loss and gain are typically better described by mixed-state open quantum dynamics, which only correspond to pure-state non-Hermitian dynamics upon a postselection of measurement outcomes. Since postselection becomes exponentially costly with particle number, we here investigate to what extent the most important example of non-Hermitian topology can survive without it: the non-Hermitian skin effect and its relationship to a bulk winding number in one spatial dimension. After defining the winding number of the Lindbladian superoperator for a quadratic fermion system, we systematically relate it to the winding number of the associated postselected non-Hermitian Hamiltonian. We prove that the two winding numbers are equal (opposite) in the absence of gain (loss), and provide a physical explanation for this relationship. When both loss and gain are present, the Lindbladian winding number typically remains quantized and non-zero, though it can change sign at a phase transition separating the loss and gain-dominated regimes. This transition, which leads to a reversal of the Lindbladian skin effect localization, is rendered invisible by postselection. We also identify a case where removing postselection induces a skin effect from otherwise topologically trivial non-Hermitian dynamics.
\end{abstract}

\maketitle

\emph{Introduction---}
The topological classification of gapped condensed matter is currently in full swing across various corners of physical parameter space~\cite{RevModPhys.89.041004, kane_TI_review}. In the context of non-interacting electrons, the theory of topological insulators and (mean-field) superconductors can be considered particularly mature. What makes this theory so successful is that it relates readily available \emph{single-particle} quantities, like a band structure’s Berry curvature, to quantized response functions such as the Hall conductivity, which are a property of the full \emph{many-body} ground state. Moreover, the bulk-boundary correspondence guarantees that a topological band structure coincides with gapless edge excitations on top of the many-body ground state, e.g. the coveted Majorana zero modes. These examples demonstrate that we must really think of topological band structures with a many-body mindset to see their most striking physics, even though they describe essentially non-interacting materials.

Topological band theory has been fruitfully generalized to non-Hermitian (NH) Hamiltonians whose spectra lie in the complex plane~\cite{kawabata_38_fold_way, PhysRevX.8.031079, Bernard2002, PhysRevLett.121.086803, Okuma_2020, PhysRevLett.124.056802, PhysRevLett.121.086803, Yang_2024, Reichel_1992, toeplitz_matrices_bottcher, PRXQuantum.4.030315, PhysRevB.103.L201114, PhysRevB.104.L161106, doi:10.1142/S1230161222500044}. Here there are different notions of gaps, the most interesting one is the \emph{point gap}, as it has no Hermitian counterpart — it denotes a region of complex energy space that is surrounded by, but does not contain, complex eigenvalues. We now have a full classification of the topological equivalence classes that emerge when a point gap is kept open in addition to various global symmetries~\cite{kawabata_38_fold_way}. At least at the single-particle level, a new NH bulk-boundary correspondence relates point-gap topology to robust boundary states without equilibrium counterpart: the most famous example is the NH skin effect in one dimension (1D), where a macroscopic number of single-particle edge states are induced by a bulk spectral winding around the point gap~\cite{zhang2022review, PhysRevLett.123.170401}. However, it is difficult to make sense of NH Hamiltonians within a many-body picture. E.g. for fermions, it is unclear how single-particle orbitals should be filled up to make a sensible many-body state. This problem is compounded by the fact that many-body NH Hamiltonians are inherently nonlocal~\cite{okuma2023review, ashida2020review, PhysRevLett.120.185301, PhysRevResearch.7.013325}.

Physically, NH Hamiltonians arise in open quantum dynamics under a postselection of measurement outcomes~\cite{Daley_2014, breuer_petruccione}. Starting with a state of $N$ particles, if measurements of the particle number are performed to guarantee no particle is lost or gained from the environment, the resulting effective dynamics is described by a NH Hamiltonian. With $N \gg 1$ however, such postselection becomes near impossible, as the chance that no quantum jump has occurred in any given time interval vanishes exponentially fast with the particle number. This means that pure states evolving under NH Hamiltonians — even those that are non-interacting — are often inadequate physical descriptions of open quantum condensed matter. To alleviate this shortcoming, we investigate to what extent postselected NH point-gap topology survives in a full Lindblad treatment of the open quantum dynamics of density matrices~\cite{lindblad1976generators,gorini1976completely,Daley_2014, CHOI1975, JAMIOLKOWSKI1972}. The analog of non-interacting NH Hamiltonians are quadratic Lindbladians with linear jump operators, which do not suffer from the aforementioned problems regarding their many-body interpretation. We first recall third quantization~\cite{Prosen_2008, Prosen_2010} which can be used to solve such quadratic systems, both bosonic and fermionic. While we focus on fermions here, the bosonic case can be treated analogously as shown in the Supplementary Material (SM)~\footnote{See the Supplementary Material for detailed derivations, including the explicit form of the steady state and relaxation modes, as well as the bosonic theory.}.

\emph{Lindbladian band structure---}
Absent interactions, the Gorini-Kossakowski-Sudarshan-Lindblad (GKSL) equation takes as input a Hermitian Hamiltonian $H$ determining the unitary part of the dynamics, and a family of jump operators $\{J_m\}_m$ that capture the interaction with an environment. The time ($\tau$) evolution of the density operator $\rho$ describing the quantum system in question is then given by 
\begin{equation} \label{eq:schrodynamicsplusquantumjumps} \mathrm{i}\frac{\mathrm{d} \rho}{\mathrm{d} \tau} = (\Hpost\rho - \rho H^{\dag}_{\textrm{post}}) + \mathrm{i} \sum_m {J_m \rho \Jdag_m}.
\end{equation}
Here we have defined the \emph{postselected} NH Hamiltonian
\begin{equation} \label{eq: postselected pre def}
    \Hpost = H - \frac{\mathrm{i}}{2}\sum_m \Jdag_m J_m
\end{equation}
that governs dynamics conditioned on the absence of quantum jump events~\cite{Daley_2014}. To determine the role of postselection in stabilizing point gap topology, we consider a U(1) charge-conserving $\Hpost$,
\begin{equation} \label{eq: postselected Ham def}
\Hpost = \sum_{ij} \mathcal{H}^{\mathrm{post}}_{ij} c^\dagger_i c_j + \emph{const.},
\end{equation}
where the single-particle NH hopping matrix $\mathcal{H}^{\mathrm{post}}_{ij}$ has translational symmetry and nontrivial point gap topology in periodic boundary conditions (PBC), and $c^\dagger_i, c_i$, $i = 1 \dots N$, are fermionic creation and annihilation operators satisfying the canonical anti-commutation relations $\{c_i,c_j\} = 0$, $\{c_i, c^\dagger_j\} = \delta_{ij}$.

To ensure $\Hpost$ is of the form of Eq.~\eqref{eq: postselected Ham def}, we restrict ourselves to a U(1) charge-conserving Hermitian Hamiltonian $H$ and two distinct sets of jump operators: losses $J_m = L_m$ and gains $J_m = G_m$ [summed over in Eqs.~\eqref{eq:schrodynamicsplusquantumjumps},~\eqref{eq: postselected pre def}], where 
\begin{equation} \label{eq: our loss gain system}
    \begin{aligned}
        &H = \sum_{ij} \mathcal{H}_{ij} c^\dagger_i c_j,\quad \mathcal{H}_{ij} = \mathcal{H}_{ji}^*, \\
        &L_m = \sum_{i} \mathcal{L}_{mi}c_i, \quad G_m = \sum_{i} \mathcal{G}_{mi}\cdag_i, 
    \end{aligned}
\end{equation}
so that, defining the Hermitian loss and gain matrices $m^{(l)} = \mathcal{L}^\dagger \mathcal{L}$ and $m^{(g)} = \mathcal{G}^\dagger \mathcal{G}$,
\begin{equation} \label{eq: Hpost matrix}
    \mathcal{H}^{\mathrm{post}}_{ij} = 
        \mathcal{H}_{ij} - \frac{\mathrm{i}}{2} m^{(l)}_{ij} + \frac{\mathrm{i}}{2} m^{(g)}_{ji}.
\end{equation}
This choice implies that the corresponding GKSL equation has \emph{weak} U(1) charge conservation symmetry~\cite{Buca_2012, ClerkWeak22}: while the total particle number $\braket{n} = \mathrm{tr} (\rho \sum_i c^\dagger_i c_i)$ is \emph{not} preserved due to the presence of the second term (the ``jump term") in Eq.~\eqref{eq:schrodynamicsplusquantumjumps}, $\rho$ does not develop quantum coherences between Hilbert space sectors of different particle number.

\begin{figure}
    \centering
    \includegraphics[width=1\linewidth]{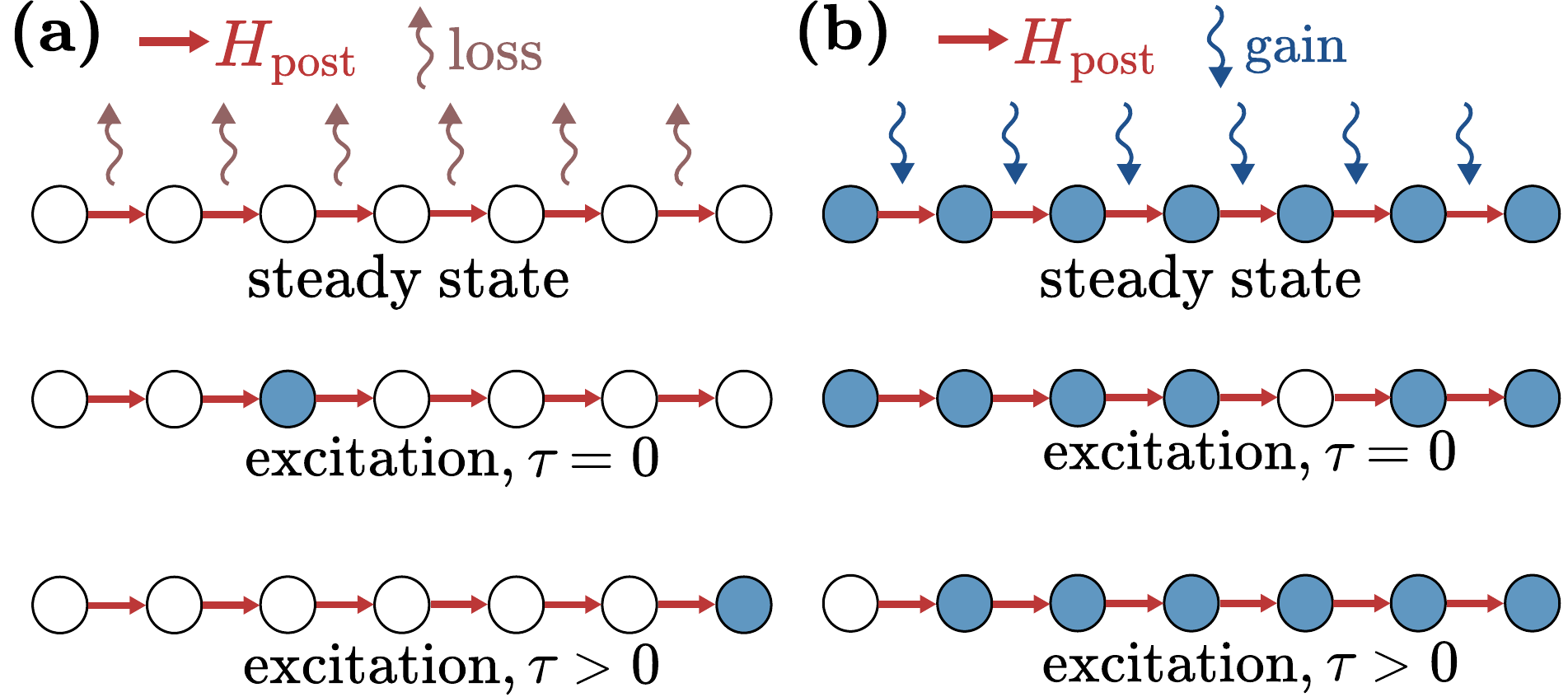}
    \caption{Dynamics of the Lindbladian `Hatano-Nelson' model [Eq.~\eqref{eq:simple_gain_loss_model_jump_operators}] under \textbf{(a)} loss only and \textbf{(b)} gain only in open boundary conditions. In both cases, the postselected NH Hamiltonian drives microscopic electrons towards the right edge (red arrows). For loss \textbf{(a)}, the non-equilibrium steady state consists of all sites empty (white circles). If we add an electron into the system (blue circle), it moves to the right. For gain \textbf{(b)}, the steady state has all sites occupied by electrons. The elementary excitations (relaxation modes) are now holes, which effectively move towards the left edge. Mathematically, this Lindbladian skin effect reversal is reflected in a Lindbladian bulk winding number that has a sign opposite to that of the postselected NH Hamiltonian.}\label{fig:loss_and_gain_chains}
\label{fig:intuition}
\end{figure}

\begin{figure*}
    \centering
    \includegraphics[width=1.0\linewidth]{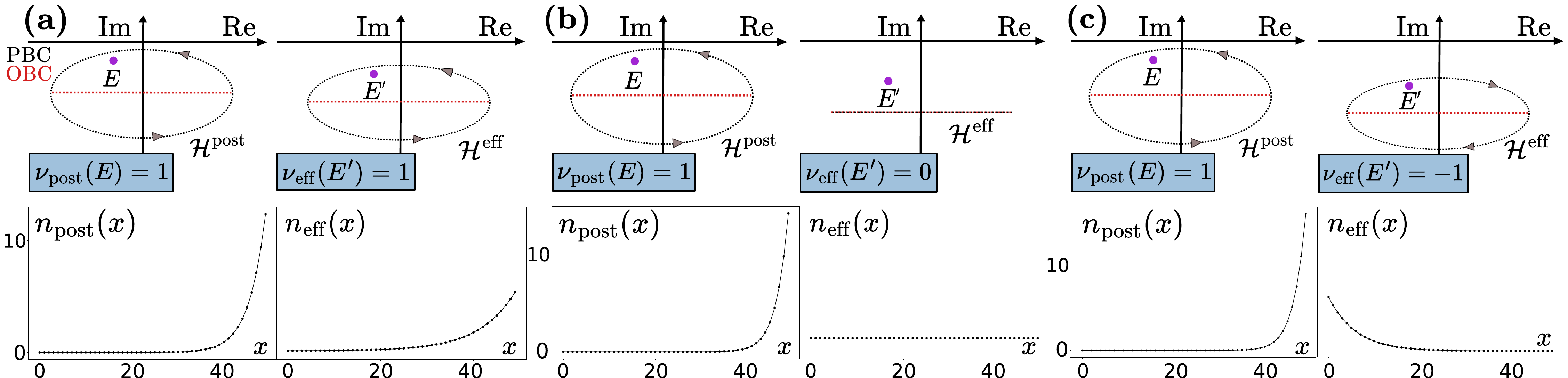}
    \caption{Comparison of postselected NH Hamiltonian $\mathcal{H}^{\mathrm{post}}(k)$ with Lindbladian band structure calculated from $\mathcal{H}^{\mathrm{eff}}(k)$ for the Lindbladian `Hatano-Nelson' model~\cite{Chaduteau_Data_2025} in Eq.~\eqref{eq:simple_gain_loss_model_jump_operators}.
    Top panels show the point gaps and their winding number [Eq.~\eqref{eq:eff_and_post_windings}] indicated by arrows for $\mathcal{H}^{\mathrm{post}}$ and $\mathcal{H}^{\mathrm{eff}}$, side by side, for both periodic and open boundary conditions. The bottom panel shows the probability density $n(x)$, summed over \emph{all} eigenstates, for $\mathcal{H}^{\mathrm{post}}$ [$n_{\mathrm{post}}(x)$] and $\mathcal{H}^{\mathrm{eff}}$ [$n_{\mathrm{eff}}(x)$], side by side.
    \textbf{(a)}~When loss is greater than gain ($\gamma_l > \gamma_g$), the winding number of $\mathcal{H}^{\mathrm{eff}}$ is the same as that of $\mathcal{H}^{\mathrm{post}}$, and a Lindbladian skin effect occurs on the same edge as the postselected NH skin effect in open boundary conditions. \textbf{(b)}~When loss equals gain ($\gamma_l = \gamma_g$), the point gap of $\mathcal{H}^{\mathrm{eff}}$ collapses, while that of $\mathcal{H}^{\mathrm{post}}$ remains open, and there is no Lindbladian skin effect. \textbf{(c)}~When gain is greater than loss ($\gamma_l < \gamma_g$), the winding number of $\mathcal{H}^{\mathrm{eff}}$ is opposite that of $\mathcal{H}^{\mathrm{post}}$, and the Lindbladian skin effect is reversed compared to the postselected NH skin effect.}
\label{fig:all_skin_effects}
\end{figure*}

To define point gap topological invariants absent postselection, we employ third quantization~\cite{Prosen_2008}. This method transforms the GKSL equation into the vectorized form
$\mathrm{i} \, \mathrm{d}\kket{\rho}/\mathrm{d}\tau = \hat{\mathcal{L}} \kket{\rho}$,
where $\hat{\mathcal{L}}$ is called the NH Lindbladian \emph{superoperator}. Here we use the symbol $\kket{\rho}$ to denote the vectorized $4^{N}$-dimensional state associated with the $2^N \times 2^N$ density matrix $\rho$. Third quantization essentially amounts to flattening the density matrix in a way that naturally respects fermion statistics. For our purposes, the main result we need is that the superoperator $\hat{\mathcal{L}}$ can be brought into the diagonal form
\begin{equation}
    \hat{\mathcal{L}} = \sum_{\alpha = 1}^{2N} \lambda_{\alpha} \hat{b}^\prime_\alpha \hat{b}_\alpha, \quad \lambda_\alpha \in \mathbb{C}, \quad \mathrm{Im}\lambda_\alpha < 0,
\end{equation}
where the operators $\hat{b}^\prime_\alpha, \hat{b}_\alpha$ are called the normal master modes (NMMs) that \emph{almost} satisfy canonical anti-commutation relations
\begin{equation}
    \{\hat{b}_\alpha, \hat{b}_\beta\} = 0, \quad \{\hat{b}_\alpha, \hat{b}^\prime_\beta \} = \delta_{\alpha \beta}, \quad \{\hat{b}^\prime_\alpha, \hat{b}^\prime_\beta\} = 0.
\end{equation}
Note that $\hat{b}^\prime_\alpha \neq \hat{b}^\dagger_\alpha$ because $\hat{\mathcal{L}}$ is not Hermitian. The non-equilibrium steady state (NESS), corresponding to $\mathrm{d} \rho/\mathrm{d} \tau= 0$, is the NMM vacuum: $\hat{b}_{\alpha} \kket{\mathrm{NESS}} = 0$. The `excitations' are the relaxation modes $\hat{b}^\prime_\alpha \hat{b}^\prime_\beta \kket{\mathrm{NESS}}$ that decay towards the steady state with a rate $e^{-\mathrm{i} (\lambda_{\alpha} + \lambda_{\beta}) \tau}$. (Here we use two $\hat{b}^\prime$ operators instead of one, because physical density matrices have even fermion parity~\cite{Note1}.)
We show in the SM~\cite{Note1} that both the `single-particle' eigenvalues $\lambda_{\alpha}$ of $\hat{\mathcal{L}}$, as well as the NMM creation operators $\hat{b}^\prime_{\alpha}$ (but \emph{not} the $\hat{b}_\alpha$), are fully determined by the spectrum and eigenvectors of the effective NH Hamiltonian~\cite{ClerkKeldysh23}
\begin{equation} \label{eq: Heff matrix}
    \mathcal{H}^{\mathrm{eff}}_{ij} = \mathcal{H}_{ij} - \frac{\mathrm{i}}{2} m^{(l)}_{ij} - \frac{\mathrm{i}}{2} m^{(g)}_{ji}.
\end{equation}
In particular, denote by $\epsilon_\alpha$ the $N$ eigenvalues of $\mathcal{H}^{\mathrm{eff}}_{ij}$, then the $\lambda_\alpha$ are given by
\begin{equation} \label{eq: specific form of the lambdas}
\begin{aligned}
    &\lambda_\alpha = \epsilon_\alpha \quad &&(\alpha = 1, \dots, N),\\
    &\lambda_\alpha = -\epsilon_{\alpha-N}^* \quad &&(\alpha = N+1, \dots, 2N).
\end{aligned}
\end{equation}
The real space profile of the NMM creation operators $\hat{b}^\prime_{\alpha}$ is exactly the same as that of the eigenvectors of $\mathcal{H}^{\mathrm{eff}}$. Another way to see that $\mathcal{H}^{\mathrm{eff}}$ governs the relaxation dynamics is the time evolution equation of the correlation matrix $\Delta_{ij} = \langle \cdag_i c_j \rangle$~\cite{mcdonald_ness_of_NH_models,PhysRevLett.123.170401}:
\begin{equation}
    \frac{\mathrm{d}\Delta}{\mathrm{d}\tau} = \mathrm{i}(\mathcal{H}^{\mathrm{eff}})^* \cdot \Delta - \mathrm{i} \Delta \cdot (\mathcal{H}^{\mathrm{eff}})^T + m^{(g)}.
\end{equation}
Since all expectation values follow from $\Delta$ via Wick's theorem, we focus on $\mathcal{H}^{\mathrm{eff}}$ rather than $\hat{\mathcal{L}}$ in the following.

In presence of translational symmetry, the complex eigenvalues of $\mathcal{H}^{\mathrm{eff}}$ yield the `band structure' of the Lindbladian superoperator and we can study its NH point gap topological invariants.
Crucially, $\mathcal{H}^{\mathrm{eff}}$ and $\mathcal{H}^{\mathrm{post}}$ are not the same: compared to $\mathcal{H}^{\mathrm{post}}$ in Eq.~\eqref{eq: Hpost matrix}, the gain term $m^{(g)}$ of $\mathcal{H}^{\mathrm{eff}}$ in Eq.~\eqref{eq: Heff matrix} comes with the opposite sign. The comparison of quadratic Lindbladian dynamics and point gap topological invariants with and without postselection then becomes an exercise of comparing $\mathcal{H}^{\mathrm{post}}$ with $\mathcal{H}^{\mathrm{eff}}$.

\emph{NH skin effect with and without postselection---}
Since $\mathcal{H}^{\mathrm{eff}}$ fully determines the NMMs but not the NESS, Lindbladian point-gap topology is a property of the evolution towards the NESS, rather than a topological characterization of the NESS itself. In the absence of gain $m^{(g)} = 0$, we can immediately deduce that the point gap topology of $\mathcal{H}^{\mathrm{post}}$ survives in the full Lindbladian dynamics -- this means that all PBC topological invariants and associated open boundary conditions (OBC) edge states are mathematically the same, even though they might have a different physical interpretation in the language of open quantum systems. In presence of gain $m^{(g)} \neq 0$, the topological \emph{classification} of $\mathcal{H}^{\mathrm{eff}}$ remains the same as that of $\mathcal{H}^{\mathrm{post}}$, but topological invariants may take on different values. 

We focus on the 1D winding number of NH symmetry class A (no symmetry) in the 38-fold way~\cite{kawabata_38_fold_way}, which for $\mathcal{H}^{\mathrm{eff}}$ is defined by
\begin{equation}\label{eq:eff_and_post_windings}
    \nu_{\mathrm{eff}} (E) = \frac{1}{2\pi\mathrm{i}}\int_{0}^{2\pi} \mathrm{d} k \, \frac{\partial}{\partial k} \log \det[\mathcal{H}^{\mathrm{eff}}(k) - E \mathbb{1}],
\end{equation}
and similarly for $\mathcal{H}^{\mathrm{post}}$. Here, $E$ is a reference complex energy inside the point gap, and $\mathcal{H}^{\mathrm{eff}}(k)$ is the momentum space NH Bloch Hamiltonian that follows via a Fourier transform from Eq.~\eqref{eq: Heff matrix}. [Note that the winding number of the second set of eigenvalues in Eq.~\eqref{eq: specific form of the lambdas}, measured around $E$, is \emph{always} equal to $\nu_{\mathrm{eff}}(-E^*)$~\cite{Note1}, which is why we do not consider it separately here.] As mentioned already, in presence of loss alone (no gain), we have $\nu_{\mathrm{eff}} = \nu_{\mathrm{post}}$. On the other hand, with gain alone (no loss), we have $\mathcal{H}^{\mathrm{eff}} = (\mathcal{H}^{\mathrm{post}})^\dagger$, from which it can be shown that $\nu_{\mathrm{eff}} = -\nu_{\mathrm{post}}$: the Lindbladian winding number and associated skin effect are \emph{opposite} that of the postselected NH Hamiltonian! Fig.~\ref{fig:intuition} illustrates this result physically for a specific $\mathcal{H}^{\mathrm{post}}$ that \emph{always} has $\nu_{\mathrm{post}} > 0$ and therefore drives electrons to the right-hand side of the system, to result in a postselected NH skin effect on the right. In presence of gain alone, the Lindbladian steady state has all lattice sites fully occupied by electrons. The Lindbladian NMM relaxation modes are therefore \emph{hole} excitations that effectively move to the left of the system, resulting in a NMM skin effect on the left and winding number $\nu_{\mathrm{eff}} < 0$. In contrast, for loss alone, the NESS is the vacuum and the NMM excitations are electrons that move to the right, confirming $\nu_{\mathrm{eff}} = \nu_{\mathrm{post}}$.
When gain and loss are both nonzero, the NESS is neither the vacuum nor the fully occupied state, and NMMs can have both electron and hole components. Surprisingly, even in this case -- as long as the point gap remains open -- \emph{all} relaxation modes still move in one direction only, irrespective of whether they add or remove electronic density. We can see this by a thought experiment: e.g. starting with gain alone, introducing a little bit of loss will not immediately close the point gap. The associated winding number $\nu_{\mathrm{eff}}$ remains quantized and cannot change continuously, meaning that there is still a skin effect in OBC at only one end of the system. Numerical simulations using the model below confirm that in this case \emph{all} perturbations on top of the steady state still move only towards the same end of the system~\cite{Note1,Chaduteau_Data_2025}. Only when gain and loss are precisely balanced is there a point gap closing at which the winding number and skin effect switches sides. This phase transition between different Lindbladian point-gap topological phases is invisible in the spectrum of $\mathcal{H}^{\mathrm{post}}$, whose point gap remains open. Knowledge of the Lindbladian winding number therefore imposes a powerful constraint on the relaxation dynamics towards the steady state (but not on the steady state itself~\cite{PhysRevLett.124.040401}), even when no interpretation as simple as that in Fig.~\ref{fig:intuition} is available. This result is model-independent and only relies on $\nu_{\mathrm{post}}$ remaining non-zero for all choices of gain and loss.

\emph{Lindbladian `Hatano-Nelson' model---}
We confirm our results in a simple Lindbladian model with a postselected skin effect that allows us to tune between gain and loss contributions. Consider an open quantum system with reciprocal hopping as well as gain and loss terms for each site, so that the index of the jump operators $L_m$ and $G_m$ becomes the same as position $m = i$, $i=1\dots L$:
\begin{equation}\label{eq:simple_gain_loss_model_jump_operators}
\begin{aligned}
    &H = t\sum_i (\cdag_{i+1}c_i + \cdag_i c_{i+1}), \\
    &L_i = \sqrt{\gamma_l}(c_i - \mathrm{i}c_{i+1}), \quad
    G_i = \sqrt{\gamma_g}(\cdag_i - \mathrm{i}\cdag_{i+1}), 
\end{aligned}
\end{equation}
where $t, \gamma_l, \gamma_g \geq 0$ and PBC are implemented by setting $c_{L+1} 
\equiv c_1$. From these we obtain
\begin{equation}\label{eq:simple_gainloss_model_hamiltonians}
\begin{aligned}
  \mathcal{H}^{\mathrm{post}}(k) &= 2t\cos k + \mathrm{i}\gamma\sin k -\mathrm{i}\delta, \\
  \mathcal{H}^{\mathrm{eff}}(k) &= 2t\cos k + \mathrm{i}\delta\sin k -\mathrm{i}\gamma,
\end{aligned}
\end{equation}
where $\gamma=\gamma_l+\gamma_g, \; \delta = \gamma_l-\gamma_g$. These expressions match that of the Hatano-Nelson model~\cite{original_hatano_nelson, Okuma_2020}, up to an imaginary shift that in $\mathcal{H}^{\mathrm{eff}}(k)$ guarantees that all relaxation modes decay towards the steady state~\cite{mcdonald_ness_of_NH_models}.
We plot the spectrum and OBC mode localization for both NH Hamiltonians in Fig.~\ref{fig:all_skin_effects}. The spectra show that the size of the point gap of $\mathcal{H}^{\mathrm{eff}}$ depends on the imbalance $\delta = \gamma_l-\gamma_g$ between loss and gain. Tuning to the critical point $\gamma_l=\gamma_g$, the spectrum collapses to a line [Fig.~\ref{fig:all_skin_effects}(b)]. The point gap in $\mathcal{H}^{\mathrm{eff}}$ closes when $\gamma_l>\gamma_g$ changes to $\gamma_l<\gamma_g$, and the observed Lindbladian skin effect changes direction (concurrent with a sign change of the winding number). At the same time, the point gap in $\mathcal{H}^{\mathrm{post}}$ remains open, and its associated NH skin effect retains the same direction.

Interestingly, it is also possible to have a Lindbladian skin effect without a NH skin effect in the postselected Hamiltonian~\cite{PhysRevLett.123.170401}. To see this, it suffices to replace the gain operators in Eq.~\eqref{eq:simple_gain_loss_model_jump_operators} by $G_i = \sqrt{\gamma_g}(\cdag_i + \mathrm{i}\cdag_{i+1})$. This has the effect of flipping $\gamma \leftrightarrow \delta$ in Eq.~\eqref{eq:simple_gainloss_model_hamiltonians}. Then, when $\gamma_g=\gamma_l$, the point gap of $\mathcal{H}^{\mathrm{post}}$ vanishes, while that of $\mathcal{H}^{\mathrm{eff}}$ remains open and exhibits a nontrivial winding number and an associated Lindbladian skin effect (see also Ref.~\cite{PhysRevLett.123.170401}).

\emph{Discussion---}
Lindbladian superoperators with nontrivial point gap topology have previously been studied in the past~\cite{PhysRevResearch.4.023160, Zhou_2022, PhysRevLett.127.070402, PhysRevB.108.054313, PhysRevLett.124.040401, Niu_2024, PhysRevResearch.6.L032067, PhysRevB.111.L060301, yang2025extended, PhysRevA.108.032214, PhysRevA.106.L011501, PhysRevA.106.032216, PhysRevLett.124.196401, PhysRevA.101.062112, PhysRevB.111.024303, Barthel_2022, soares2025dissipative, Wanjura2020, PhysRevLett.127.213601, 10.21468/SciPostPhys.15.4.173, Roccati2024}, and shown to exhibit Lindbladian skin effects, also dubbed Liouvillian skin effects (see also Refs.~\onlinecite{gideon2023_anomalousl_relaxations_NH,Niu_2024, Liouvillian_line_gap_topology, xiao2024topology, longhi_nhse_dissip, xiao2023local, PhysRevB.109.014313, PhysRevB.108.155114, PhysRevB.110.045440, yang2025quantumdynamicalsignaturesnonhermitian}). 
In addition, there have been initial explorations of the topological aspects of non-equilibrium steady states~\cite{bardyn2013topology, PhysRevB.91.165140, PhysRevX.8.011035, PhysRevResearch.5.023004}. What sets our paper apart is that we propose a systematic definition of the NH band structure associated with a quadratic Lindblad problem, and directly compare its NH topological invariants with those of the corresponding postselected NH Bloch Hamiltonian. We have shown that in the case of quadratic Lindbladians with a weak $U(1)$ symmetry, while the 1D winding number remains a topological invariant without postselection, its value and physical interpretation in terms of a Lindbladian skin effect can change completely. A natural future direction is to generalize our approach to higher dimensions, other NH symmetry classes, and their respective topological invariants. Another promising idea is to determine the constraints that nontrivial Lindbladian point-gap topology imposes on the non-equilibrium steady state itself.
Finally, as we have shown here, the topological gain-loss phase transition can become invisible after postselection, but it does not have to. One should compare this transition to the recent examples of measurement-induced phase transitions in the entanglement entropy that are \emph{only} visible after postselection~\cite{Li_2025}.

\begin{acknowledgments}
We thank Julia Hannukainen for a critical reading of the manuscript and insightful discussions. 
We also thank Peru d'Ornellas, Eva-Maria Graefe, Ryan Barnett, Nicolas Regnault, Titus Neupert, and Alessandro Romito for insightful discussions.
A.C. acknowledges support from Imperial College London via  a President’s PhD Scholarship. This work was supported by a UKRI Future Leaders Fellowship MR/Y017331/1.
\end{acknowledgments}

\bibliography{refs}

\end{document}

\title{Supplemental Material for\\ ``Lindbladian versus Postselected Non-Hermitian Topology"}

\author{Alexandre Chaduteau}
\affiliation{Blackett Laboratory, Imperial College London, London SW7 2AZ, United Kingdom}

\author{Derek K. K. Lee}
\affiliation{Blackett Laboratory, Imperial College London, London SW7 2AZ, United Kingdom}

\author{Frank Schindler}
\affiliation{Blackett Laboratory, Imperial College London, London SW7 2AZ, United Kingdom}

\maketitle

\tableofcontents

\section{Lindbladian open quantum dynamics}
We first quickly review Lindbladian~\cite{lindblad1976generators, gorini1976completely} dynamics of open quantum systems. For a much more pedagogical introduction, see e.g. Ref.s~\cite{breuer_petruccione, Daley_2014}. Consider a density operator $\rho$ describing the state of our system. For example, if our quantum system is in a pure state described by $|\psi\rangle$, then $\rho = |\psi\rangle \langle \psi |$. With this definition, the Lindblad (or GKSL) equation~\cite{lindblad1976generators, gorini1976completely, Daley_2014} is a dynamical equation of the density matrix that can be derived under a series of assumptions, the most prominent being the Born approximation, Markovianity, and the rotating-wave approximation~\cite{breuer_petruccione}. It can be expressed as
\begin{equation}\label{eq:lindblad_dynamics}
\mathrm{i} \frac{\mathrm{d} \rho}{\mathrm{d} \tau} := \mathcal{L}\rho = \left[ H, \rho \right] + \mathrm{i} \sum_m \left[ {J_m \rho \Jdag_m - \frac{1}{2}\left\{ \Jdag_m J_m, \rho \right\} } \right],
\end{equation}
where $H$ is a Hermitian Hamiltonian and $J_m$ are “jump operators", which are operators on our Fock space that describe interactions with the environment. For our purposes studying one-dimensional systems (with $N$ lattice sites), these are labelled by an integer $m$, and there can be any number\footnote{Although in the models we present here, we have as many jumps as lattice sites, i.e. $m \in \{1, \ldots N\}$ [see Sec.~\ref{sec:explicit_example}].} of such jumps~\cite{breuer_petruccione}. When all jumps are zero, we recover Schrödinger/von Neumann dynamics for closed quantum systems. Also note that we adopt the convention whereby decay frequencies are directly incorporated in the definition of the jump operators. The Lindbladian $\mathcal{L}$ is a “super"-operator: $\mathcal{L}: \rho(\mathcal{H}) \rightarrow \rho(\mathcal{H})$ where $\rho(\mathcal{H})$ is the space of all density matrices on the Hilbert space $\mathcal{H}$.

One can derive a conditional non-Hermitian (NH) Hamiltonian from this: let 
\begin{equation}
\Hpost = H -\frac{\im}{2}\sum_m {J^{\dagger}_m J_m},
\end{equation}
then the Lindblad equation becomes
\begin{equation} \label{eq:schrodynamicsplusquantumjumps}
\frac{\mathrm{d} \rho}{\mathrm{d} \tau} = -\im(\Hpost\rho - \rho H^{\dag}_{\textrm{post}}) + \sum_m {J_m \rho \Jdag_m}.
\end{equation}
The quantum jump term $Q = \sum_m {J_m \rho \Jdag_m}$ is the only difference to purely Schrödinger dynamics on a NH Hamiltonian $\Hpost$. We refer to dynamics without the $Q$ term as \emph{postselected} dynamics, as it corresponds to selecting system trajectories where no quantum jumps between different particle-number sectors occurred~\cite{breuer_petruccione} (see also main text).
\section{Third quantization of open quantum systems}\label{sec:third_quantization}
Our aim is to understand the point gap topology of $\Hpost$ in the general Lindbladian framework of Eq.~\eqref{eq:schrodynamicsplusquantumjumps}. We will exclusively work with non-interacting fermion systems and linear jump operators. For this case, the (exponentially complex) GKSL time evolution can be solved efficiently using Prosen's third quantization formalism~\cite{Prosen_2008}, which we review in the following. 

In this section, to distinguish all types of indices, we use
\begin{itemize}
\item{Latin indices $i, j, l= 1 \dots N$ for real-space, complex fermions,}
\item{Greek indices $\alpha, \beta, \ldots = 1 \dots 2N$ for Majorana fermions and super-fermions,} 
\item{$k$ for momenta in the one-dimensional Brillouin Zone $\frac{2\pi}{N}\{0, \dots, N-1\}$.}
\end{itemize}
\subsection{Review of formalism}
Third quantization is a method of explicit solution for \emph{quadratic} Hamiltonians with \emph{linear} jump operators. For a system of $N$ complex fermions with annihilation operators $c_i, \; i \in \{1, \ldots N\}$ (as in the main text), we can define Majorana operators $\gamma_{\alpha}, \; \alpha \in \{1, \ldots 2N\}$ via
\begin{equation}
    c_i = \frac{1}{2}(\gamma_{2i-1} - \im \gamma_{2i}).
\end{equation}
These Majoranas satisfy $\{\gamma_{\alpha}, \gamma_{\beta}\} = 2\delta_{\alpha \beta}, \; \alpha, \beta = 1, \ldots 2N$. With this, we decompose our Lindblad dynamics' Hamiltonian and jump operators into Majoranas via
\begin{equation} \label{eq:majorana_definitions}
    H = \sum_{\alpha,\beta=1}^{2N}\gamma_{\alpha} H_{\alpha \beta}\gamma_{\beta}, \qquad
    J_{m} = \sum_{\alpha=1}^{2N}J_{m, \alpha}\gamma_{\alpha}.
\end{equation}
The Majoranas' anti-commutation relations ensure we may choose the Majorana Hamiltonian $H$ to be anti-symmetric i.e. $H^T = -H$, up to a constant energy shift. Moreover, Hermiticity of $H = H^\dagger$ guarantees $H \in \mathbb{R}^{2N \times 2N}$.
The usual density operators/matrices $\rho$ of which we consider the Lindbladian evolution~\eqref{eq:lindblad_dynamics} can be added together and multiplied by scalars; they live in the vector space $\rho({\mathcal{H}})$. We then define the ($2^N \times 2^N = 4^N$)-dimensional Fock-Liouville space $\mathcal{K} \cong \mathcal{H} \otimes \mathcal{H}^*$~\cite{CHOI1975, JAMIOLKOWSKI1972}, which consists of flattened/“vectorized" density operators, i.e. we map $\rho \rightarrow \kket{\rho} \in \mathcal{K}$. The space $\rho(\mathcal{H})$ induces a natural inner product in $\mathcal{K}$ via 
\begin{equation}
\langle\!\langle{x}|y\rangle\!\rangle := \frac{1}{2^N}\mathrm{Tr}(x^{\dagger}y).
\end{equation}
We also note linearity, for $a,b \in \mathbb{C}$:
\begin{equation} \label{eq: linearity}
    \kket{a x + b y} = a \kket{x} + b \kket{y}.
\end{equation}
The use of this formalism is to define the Lindbladian $\mathcal{L}$ as an operator acting on $\mathcal{K}$, allowing to express it as a matrix, independent of $\rho$ in~\eqref{eq:lindblad_dynamics}.
As per Ref.~\cite{Prosen_2008}, the space $\mathcal{K}$ has a canonical basis $\{\kket{P_{\bs{n}}}\}_{\bs{n}}$ where
\begin{equation}\label{eq:canonical_liouville_basis}
    P_{n_1, n_2, \ldots, n_{2N}} := \gamma_{1}^{n_1}\gamma_{2}^{n_2}\ldots\gamma_{2N}^{n_{2N}}, \quad n_{\alpha} \in \{0, 1\}, \quad \alpha \in \{1, \ldots 2N\}.
\end{equation}
This can be seen as spanning just another Fock space: we define $2N$ so-called \emph{super-fermion}~\cite{Prosen_2008} annihilation and creation operators $(\chat_\alpha,\chatdag_\alpha)$ respectively by 
\begin{equation}
    \hat{c}_\alpha\kket{P_{\bs{n}}} = \delta_{n_\alpha, 1}\kket{\gamma_\alpha P_{\bs{n}}}, \qquad \hat{c}_\alpha^{\dagger}\kket{P_{\bs{n}}} = \delta_{n_\alpha, 0}\kket{\gamma_\alpha P_{\bs{n}}}.
\end{equation}
These super-fermions obey the usual canonical anti-commutation relations $\{\chat_\alpha,\chat_\beta\} = 0$, $\{\chat_\alpha, \chatdag_\beta\} = \delta_{\alpha\beta}$, as can be seen by acting on an arbitrary basis element $\kket{P_{\bs{n}}}$. We put hats on these super-fermion operators, to distinguish them from (regular) fermion operators $c_i, \: i \in \{1,\ldots N\}$ on the original Hilbert space.

\subsection{Lindbladian super-operator}\label{sec:main_lindbladian_derivation}
We can express the action of the Lindbladian $\mathcal{L}$ in terms of the super-fermions. We denote the corresponding third-quantized super-operator by $\hat{\mathcal{L}}$. Firstly, the free/unitary part of Lindblad dynamics maps to a ket in $\mathcal{K}$ as
\begin{equation}
    \mathcal{L}_0\rho \rightarrow \kket{\mathcal{L}_0\rho} = \kket{\left[ H, \rho \right]}.
\end{equation} 
Using the definition of $H$ in Eq.~\eqref{eq:majorana_definitions}, we have
\begin{equation}
    \kket{\left[ H, \rho \right]} = \sum_{\alpha, \beta=1}^{2N} H_{\alpha \beta} \kket{[\gamma_\alpha \gamma_\beta, \rho]} = \sum_{\alpha \beta} H_{\alpha \beta} \left(\kket{\gamma_\alpha \gamma_\beta \rho} - \kket{\rho \gamma_\alpha \gamma_\beta} \right).
\end{equation}
We first calculate this commutator for an arbitrary matrix element $\kket{P_{\bs{n}}}$:
\begin{equation}\label{eq:lie_derivative}
\begin{aligned}
  \kket{\gamma_\alpha\gamma_\beta P_{\bs{n}}}-\kket{P_{\bs{n}}\gamma_\alpha\gamma_\beta}
    &= 2(\delta_{n_\alpha, 1}\delta_{n_\beta, 0} + \delta_{n_\alpha, 0}\delta_{n_\beta, 1})\kket{\gamma_\alpha\gamma_\beta P_{\bs{n}}}\\
    &=2(\hat{c}_\alpha\hat{c}_\beta^{\dagger} + \hat{c}_\alpha^{\dagger}\hat{c}_\beta)\kket{P_{\bs{n}}}\\
    &=2(\hat{c}_\alpha^{\dagger}\hat{c}_\beta -\hat{c}_\beta^{\dagger}\hat{c}_\alpha )\kket{P_{\bs{n}}},
\end{aligned}
\end{equation}
where the first equality follows from anti-commutation relations via 
\begin{equation}
\begin{aligned}
    P_{\bs{n}}\gamma_\alpha\gamma_\beta = (-1)^{|\bs{n}| - n_\alpha + |\bs{n}| - n_\beta}\gamma_\alpha \gamma_\beta P_{\bs{n}} \\
    \implies  \left[ \gamma_\alpha \gamma_\beta, P_{\bs{n}} \right]&= (1-(-1)^{-n_\alpha - n_\beta})\gamma_\alpha \gamma_\beta P_{\bs{n}}\\
    &= 2(\delta_{n_\alpha, 1}\delta_{n_\beta, 0} + \delta_{n_\alpha, 0}\delta_{n_\beta, 1})\gamma_\alpha\gamma_\beta P_{\bs{n}}.
\end{aligned}
\end{equation}
By linearity, we can extend~\eqref{eq:lie_derivative} to any element of $\mathcal{K}$, giving
\begin{equation}
    \hat{\mathcal{L}}_0= 2\sum_{\alpha, \beta}H_{\alpha \beta}(\hat{c}^{\dagger}_\alpha\hat{c}_\beta - \hat{c}^{\dagger}_\beta\hat{c}_\alpha) = 4\sum_{\alpha, \beta}\hat{c}_\alpha^{\dagger}H_{\alpha \beta}\hat{c}_\beta := 4\bs{\hat{c}}^{\dagger}\cdot H\bs{\hat{c}},
\end{equation}
where in the penultimate equality we used anti-symmetry of the Majorana Hamiltonian.

Now consider the non-unitary part of~\eqref{eq:lindblad_dynamics}. For conciseness, we only treat the case of loss jump operators $J_m=L_m$, but the addition of gain operators $G_m$ is straightforward. Let
\begin{equation}
    -\mathrm{i}\mathcal{L}_m \rho := L_m \rho L_m ^{\dagger} - \frac{1}{2}\{L_m ^{\dagger} L_m ,  \rho\}=\sum_{\alpha, \beta}L_{m, \alpha}L_{m, \beta}^{*}\mathcal{L}_{\alpha \beta}\rho,
\end{equation}
where $-\mathrm{i}\mathcal{L}_{\alpha \beta}\rho := \gamma_\alpha\rho\gamma_\beta - \frac{1}{2}\gamma_\beta \gamma_\alpha \rho -\frac{1}{2} \rho\gamma_\beta\gamma_\alpha$.
Using anti-commutation relations again, one finds that 
\begin{equation}\label{eq:first_step_deriving_jump_lindbladian}
    -\mathrm{i}\hat{\mathcal{L}}_{\alpha \beta} \kket{P_{\bs{n}}} = \left[(-1)^{|\bs{n}| + n_\beta}\kket{\gamma_\alpha\gamma_\beta P_{\bs{n}}} - \frac{1}{2}\kket{\gamma_\beta\gamma_\alpha P_{\bs{n}}} -\frac{1}{2}(-1)^{n_\alpha + n_\beta}\kket{\gamma_\beta\gamma_\alpha P_{\bs{n}}}\right].
\end{equation}
Using the identities
\begin{equation}\label{eq:liouville_basis_identities}
\begin{aligned}
    \kket{\gamma_\alpha P_{\bs{n}}} &= (\hat{c}_\alpha^{\dagger}+\hat{c}_\alpha)\kket{P_{\bs{n}}}\\
    (-1)^{n_\alpha}\kket{\gamma_\alpha P_{\bs{n}}} &= (\hat{c}_\alpha^{\dagger}-\hat{c}_\alpha)\kket{P_{\bs{n}}} \\
    (-1)^{|\bs{n}|}\kket{P_{\bs{n}}} &= e^{\mathrm{i}\pi\hat{\mathcal{N}}}\kket{P_{\bs{n}}},
\end{aligned}
\end{equation}
where $\hat{\mathcal{N}} = \sum_\alpha \chatdag_\alpha \chat_\alpha$ is the super-fermion number~\cite{Prosen_2008}, one obtains
\begin{equation}
\begin{split}
    -\mathrm{i}\hat{\mathcal{L}}_{\alpha \beta} \kket{P_{\bs{n}}} =(\hat{c}_\alpha^{\dagger} + \hat{c}_\alpha)(\hat{c}_\beta^{\dagger} -\hat{c}_\beta)e^{\mathrm{i}\pi\hat{\mathcal{N}}}\kket{P_{\bs{n}}} \\
    - \frac{1}{2}(\hat{c}_\beta^{\dagger} + \hat{c}_\beta)(\hat{c}_\alpha^{\dagger} +\hat{c}_\alpha)\kket{P_{\bs{n}}} \\
    -\frac{1}{2}(\hat{c}_\beta^{\dagger} - \hat{c}_\beta)(\hat{c}_\alpha^{\dagger} - \hat{c}_\alpha)\kket{P_{\bs{n}}}\\
    - \delta_{\alpha \beta}\kket{P_{\bs{n}}},
\end{split}
\end{equation}
which yields
\begin{equation}\label{eq:deriving_lindblad_part_1}
\begin{split}
    -\mathrm{i}\hat{\mathcal{L}}_{\alpha \beta} = \hat{c}_\alpha^{\dagger}\hat{c}_\beta^{\dagger}\left[1+ e^{\mathrm{i}\pi\hat{\mathcal{N}}}\right] + \hat{c}_\alpha\hat{c}_\beta\left[ 1 - e^{\mathrm{i}\pi\hat{\mathcal{N}}}\right] + \hat{c}_\alpha\hat{c}_\beta^{\dagger}e^{\mathrm{i}\pi\hat{\mathcal{N}}} -  \hat{c}_\alpha^{\dagger}\hat{c}_\beta e^{\mathrm{i}\pi\hat{\mathcal{N}}} \\ 
    - \frac{1}{2}\hat{c}_\beta^{\dagger}\hat{c}_\alpha - \frac{1}{2}\hat{c}_\beta\hat{c}_\alpha^{\dagger} + \frac{1}{2}\hat{c}_\beta^{\dagger}\hat{c}_\alpha + \frac{1}{2}\hat{c}_\beta\hat{c}_\alpha^{\dagger} - \delta_{\alpha \beta},
\end{split}
\end{equation}
which after rearranging gives
\begin{equation}\label{eq:deriving_lindblad_part_2}
\begin{split}
    -\mathrm{i}\hat{\mathcal{L}}_{\alpha \beta} = \frac{1}{2}(1+e^{\mathrm{i}\pi\hat{\mathcal{N}}})(2\hat{c}_\alpha^{\dagger}\hat{c}_\beta^{\dagger}-\hat{c}_\alpha^{\dagger}\hat{c}_\beta-\hat{c}_\beta^{\dagger}\hat{c}_\alpha)\\
    +\frac{1}{2}(1- e^{\mathrm{i}\pi\hat{\mathcal{N}}})(2\hat{c}_\alpha\hat{c}_\beta-\hat{c}_\alpha\hat{c}_\beta^{\dagger}-\hat{c}_\beta\hat{c}_\alpha^{\dagger}).
\end{split}
\end{equation}
For a discussion of this equation and conservation of super-fermion parity, see~\cite{Prosen_2008}. We focus on the positive ($+$) super-fermion parity sector ($e^{\mathrm{i}\pi\hat{\mathcal{N}}} = 1$) so that the total Lindblad super-operator is
\begin{equation}
    \Lplus =\hat{\mathcal{L}}_0 + \sum_{m}\hat{\mathcal{L}}^+_m
 = 4\bs{\hat{c}}^{\dagger}\cdot H\bs{\hat{c}} + \mathrm{i}\sum_{m}\sum_{\alpha, \beta}L_{m, \alpha}L_{m, \beta}^{*}(2\hat{c}_\alpha^{\dagger}\hat{c}_\beta^{\dagger}-\hat{c}_\alpha^{\dagger}\hat{c}_\beta-\hat{c}_\beta^{\dagger}\hat{c}_\alpha). 
\end{equation}
We now define the $2N\times2N$ matrix
\begin{equation}\label{eq:M_matrix_definition}
    M_{\alpha \beta} = \sum_m L_{m, \alpha}^*L_{m, \beta},
\end{equation}
which more simply\footnote{This is a different convention to Ref.~\cite{Prosen_2008} who define $M = L^TL^*$.} is $M = L^\dagger L$. Note that $M^*_{\alpha\beta} = M^T_{\alpha\beta}$. With this, one obtains 
\begin{equation}
\begin{aligned}
    \Lplus &= 4\bs{\hat{c}}^{\dagger}H\bs{\hat{c}} + \mathrm{i}\sum_{\alpha, \beta} M_{\beta\alpha}(2\hat{c}_\alpha^{\dagger}\hat{c}_\beta^{\dagger}-\hat{c}_\alpha^{\dagger}\hat{c}_\beta-\hat{c}_\beta^{\dagger}\hat{c}_\alpha) \\ 
    &= \bs{\hat{c}}^{\dagger}\cdot (4H -\mathrm{i} M -\mathrm{i} M^T)\bs{\hat{c}}-\mathrm{i}\bs{\hat{c}}^{\dagger}\cdot (M-M^T)\bs{\hat{c}}^{\dagger}\\
    &= \frac{1}{2}\left[\bs{\hat{c}}^{\dagger}\cdot Z\bs{\hat{c}}-\bs{\hat{c}}\cdot Z^T\bs{\hat{c}}^{\dagger}+\bs{\hat{c}}^{\dagger}\cdot 2Y\bs{\hat{c}}^{\dagger}\right] + \frac{1}{2}\mathrm{Tr}Z\mathbb{1},
\end{aligned}
\end{equation}
where we defined 
\begin{equation}
\begin{aligned}
    Z_{\alpha\beta} &= 4H_{\alpha\beta} - 2\mathrm{i}\mathrm{Re}M_{\alpha\beta}, \quad \\
    Y_{\alpha\beta} &= 2\mathrm{Im}M_{\alpha\beta}.
\end{aligned}
\end{equation}
[Note that $\mathrm{Tr}Z = -2\mathrm{iTr}M$ because $H$ is anti-symmetric.] By analogy with the usual Bogoliubov-de Gennes formalism for superconductors, we may elegantly define
\begin{equation}
    \hat{\mathcal{L}}_{+} = \frac{1}{2}(\underline{\chat}^\dagger, \; \underline{\hat{c}})
    \LBDG
    \begin{pmatrix}
        \underline{\chat}\\
        \underline{\chat}^\dagger
    \end{pmatrix} - \mathrm{iTr}M\mathbb{1}:= \frac{1}{2}\Psi^\dagger \LBDG \Psi - \mathrm{iTr}M\mathbb{1},
\end{equation}
where we defined~\cite{Prosen_2008, PhysRevLett.124.040401} the upper triangular matrix
\begin{equation}\label{eq:real_space_bdg_matrix}
    \LBDG = \begin{pmatrix}
        Z & 2Y \\
        0 & -Z^T
    \end{pmatrix}.
\end{equation}
This equation has the same form as the usual BdG Hamiltonian for superconductors, except there is no bottom-diagonal superconducting term $\Delta^*$ due to the absence of $\hat{c}\hat{c}$ terms in the (positive super-fermion parity) Lindbladian. Thus, unlike BdG Hamiltonians, this system does not have an inbuilt particle-hole symmetry (PHS). 
\subsection{Diagonalization of the real-space BdG matrix} Eq.~\eqref{eq:real_space_bdg_matrix} is traceless. In fact one can easily show
\begin{equation}
    \mathrm{det}(\LBDG -\lambda\mathbb{1}) = \mathrm{det}(Z-\lambda\mathbb{1)}\mathrm{det}(-Z-\lambda\mathbb{1}),
\end{equation}
meaning eigenvalues come in pairs $\pm \lambda$, for $\lambda$ an eigenvalue of $Z$. Define the matrix $S$ such that 
\begin{equation}\label{eq:change_basis_rid_Y_realspace}
    S^{-1}\LBDG S = \begin{pmatrix}
        Z & 0 \\
        0 & -Z^T
    \end{pmatrix} := \tilde{\mathcal{L}}_+.
\end{equation}
We can surmise that this matrix takes the form 
\begin{equation}
    S = \begin{pmatrix}
        1 & X\\
        0 & 1
    \end{pmatrix} \implies ZX + XZ^T = -2Y,
\end{equation}
in order for Eq.~\eqref{eq:change_basis_rid_Y_realspace} to hold. This is known as a continuous Lyapunov equation~\cite{Prosen_2010}. It defines the matrix $X$, which contains information about $Y$.
We also define diagonalizing matrices $V$ and $W$ such that
\begin{equation}
    \begin{aligned}
        Z = V^{-1}\Lambda_z V, \\
        \tilde{\mathcal{L}} = W^{-1}\Lambda W,
    \end{aligned}
\end{equation}
where $\Lambda_z$ is the diagonalized $Z$, and
\begin{equation}
    \Lambda = \begin{pmatrix}
        \Lambda_z & 0 \\
        0 & -\Lambda_z
    \end{pmatrix}
\end{equation}
is diagonal and consists of pairs of eigenvalues of $Z$ and $-Z^T$, as discussed above. Then we see that
\begin{equation}
    W = \begin{pmatrix}
        V & 0 \\
        0 & (V^{-1})^T
    \end{pmatrix}.
\end{equation}
Using the above, we see that $\hat{\mathcal{L}}_+$ is diagonalized by the combination of $S$ and $W$, i.e.
\begin{equation}
    \hat{\mathcal{L}}_+ = \frac{1}{2}\Psi^\dagger SW^{-1}\Lambda W S^{-1}\Psi  -\mathrm{iTr}M\mathbb{1}.
\end{equation}
Let 
\begin{equation}
    P = WS^{-1} = \begin{pmatrix}
        V & 0 \\
        0 & (V^{-1})^T
    \end{pmatrix}
    \cdot
    \begin{pmatrix}
        1 & X \\
        0 & 1
    \end{pmatrix} = \begin{pmatrix}
        V & -VX \\
        0 & (V^{-1})^T
    \end{pmatrix}.
\end{equation}
Then we can define new Nambu spinors
\begin{equation}
    \begin{pmatrix}
        \underline{\hat{b}}\\
        \underline{\hat{b}}^\prime
    \end{pmatrix} = P \begin{pmatrix}
        \underline{\chat}\\
        \underline{\chat}^\dagger
    \end{pmatrix}.
\end{equation}
This yields
\begin{equation}
    \begin{aligned}
        \hat{b}_\alpha &= \sum_\beta \left[V_{\alpha\beta}\chat_\beta - (VX)_{\alpha\beta}\chatdag_\beta\right], \\
        \hat{b}^\prime_\alpha &= \sum_\beta (V^{-1})^T_{\alpha\beta}\chatdag_\beta.
    \end{aligned}
\end{equation}
We also have 
\begin{equation}
    P^{-1} = \begin{pmatrix}
        V^{-1} & XV^T \\
        0 & V^T
    \end{pmatrix}
\end{equation}
and hence (sum over $\beta$ implicit)
\begin{equation}
    \begin{pmatrix}
        \underline{\chatdag}, \underline{\chat}
    \end{pmatrix}\cdot P^{-1} = \begin{pmatrix}
        \chatdag_\beta (V^{-1})_{\beta\alpha}, \quad\chatdag_\beta(XV^T)_{\beta\alpha} + \chat_\beta (V^T)_{\beta\alpha}
    \end{pmatrix}_\alpha = \begin{pmatrix}
        \hat{b}^\prime_\alpha, \quad \hat{b}_\alpha
    \end{pmatrix}_\alpha,
\end{equation}
assuming $X^T = -X$.
Thus we get
\begin{equation}
\begin{aligned}
    \Lplus &= \frac{1}{2}\begin{pmatrix}
        \underline{\hat{b}}^\prime, \quad \underline{\hat{b}}
    \end{pmatrix}
    \begin{pmatrix}
        \Lambda_z & 0\\
        0 & -\Lambda_z
    \end{pmatrix}
    \begin{pmatrix}
        \underline{\hat{b}} \\
        \underline{\hat{b}}^\prime
    \end{pmatrix} -\mathrm{iTr}M\mathbb{1}\\
    &= \underline{\hat{b}}^\prime \Lambda_z         \underline{\hat{b}} -\frac{1}{2}(\sum_\alpha \lambda_\alpha) \mathbb{1} -\mathrm{iTr}M\mathbb{1} \\
    &= \underline{\hat{b}}^\prime \Lambda_z         \underline{\hat{b}}.
\end{aligned}
\end{equation}
We used the fact that the $\hat{b}, \hat{b}^\prime$ operators satisfy almost-canonical commutation relations~\cite{Prosen_2008}. These operators are the normal master modes discussed in the main text [see Eq. (6) there].

As explained in the main text, single-particle eigenmodes of the Lindbladian are $\hat{b}^{\prime}_\alpha \kket{\mathrm{NESS}}$, where $\kket{\mathrm{NESS}}$ is the third-quantized/vectorized form of the non-equilibrium steady-state, defined as $\hat{b}_\alpha \kket{\mathrm{NESS}} = 0$ for any $\alpha$. For eigenmodes, the $\hat{b}^{\prime}_\alpha$ operators only depend on $Z$ and hence only on $\mathcal{H}^{\mathrm{eff}}$ (introduced in the main text and defined in Sec.~\ref{sec:winding_derivation}.), while the NESS also depends on the $Y$ matrix. On the other hand, single-particle eigenvalues of the Lindbladian, which dictate the rate of approach of eigenmodes to the NESS, only depend on $\mathcal{H}^{\mathrm{eff}}$. A skin effect is a statement about localization of the dynamical eigenmodes, which must occur in the $\hat{b}^{\prime}_\alpha$. While their spatial profile is indeed superposed on top of the NESS's profile, their localization is independent of it. A skin effect in the $\hat{b}^{\prime}_\alpha$ operators necessarily stems from a skin effect in $\mathcal{H}^{\mathrm{eff}}$.
\section{Momentum space}\label{sec:PBC}
All the above “Majorana" matrices (e.g. $H_{\alpha\beta},\; M_{\alpha\beta},\; Z_{\alpha\beta}\ldots$) have index $\alpha \in \{1 \ldots 2N\}$ that can be re-shuffled into $\alpha = (i,\mu)$, where as before $i \in \{1, \ldots N\}$ is a lattice site index and $\mu \in \{1, 2\}$ is an index labelling the two Majoranas at each site. Similarly $\beta = (j,\nu)$, where $j \in \{1, \ldots N\}$ and $\nu \in \{1, 2\}$. Then we can write e.g. $Z_{\alpha,\beta} := Z_{i\mu,j\nu}$. We now assume translational invariance, i.e. the Majorana matrices can be rewritten as 
\begin{equation}
Z_{i\mu,j\nu} = Z_{i-j,\mu\nu} := Z_{a,\mu\nu}, 
\end{equation}
where $a:=i-j$ is a “range index" that is assumed to take values $a \in \{-a_{\text{max}}, \ldots, a_{\text{max}}\}$ symmetric about $0$. This assumption allows for a well-defined crystal momentum $k \in \frac{2\pi}{N}\{0, \dots, N-1\}$. We Fourier transform the super-fermions via
\begin{equation}
    \chat_{i,\mu}=\frac{1}{\sqrt{N}}\sum_{k}\chat_{k,\mu}e^{-\mathrm{i}ki}.
\end{equation}
Then from the discussion surrounding Eq.~\eqref{eq:real_space_bdg_matrix}, one obtains
\begin{equation}
\begin{aligned}
    \Lplus &= \frac{1}{2}\sum_{i,j,\mu,\nu}\left[\chatdag_{i,\mu}Z_{i-j,\mu\nu}\chat_{j,\nu} - \chat_{i,\mu}Z_{j-i,\nu\mu}\chatdag_{j,\nu} + \chatdag_{i,\mu}2Y_{i-j,\mu\nu}\chatdag_{j,\nu}\right]  -\mathrm{iTr}M\mathbb{1} \\
    &= \frac{1}{2}\sum_{a,j,\mu,\nu}\left[ \chatdag_{j+a,\mu}Z_{a,\mu\nu}\chat_{j,\nu} - \chat_{j+a,\mu}Z_{-a,\nu\mu}\chatdag_{j,\nu} + \chatdag_{j+a,\mu}2Y_{a,\mu\nu}\chatdag_{j,\nu}\right] - \mathrm{iTr}M\mathbb{1}\\
    &= \frac{1}{2N}\sum_{a,j,\mu,\nu, k, q}\left[ \chatdag_{k,\mu}e^{\mathrm{i}k(j+a)}Z_{a,\mu\nu}\chat_{q,\nu}e^{-\mathrm{i}qj} - \chat_{k,\mu}e^{-\mathrm{i}k(j+a)}Z_{-a,\nu\mu}\chatdag_{q,\nu}e^{\mathrm{i}qj} + \chatdag_{k,\mu}e^{\mathrm{i}k(j+a)}2Y_{a,\mu\nu}\chatdag_{q,\nu}e^{\mathrm{i}qj}\right] -\mathrm{iTr}M\mathbb{1} \\
    &= \frac{1}{2}\sum_{k,\mu,\nu}\left[\chatdag_{k,\mu}Z(k)_{\mu\nu}\chat_{k,\nu} - \chat_{k,\mu}Z(k)_{\nu\mu}\chatdag_{k,\nu} + \chatdag_{k,\mu}2Y(k)_{\mu\nu}\chatdag_{-k,\nu}\right] - \mathrm{iTr}M\mathbb{1},
\end{aligned}
\end{equation}
where we used some delta functions, and defined Fourier-transformed quantities as e.g.
\begin{equation}\label{eq:fourier_transform_definition_lindblad}
    Z(k)_{\mu\nu} := \sum_a Z_{a,\mu\nu}e^{\mathrm{i}ka}
\end{equation}
and used symmetry of the range index $a$ to flip $a \rightarrow -a$ in the second term. As in real space, we can re-write this as
\begin{equation}
    \hat{\mathcal{L}}_+ = \frac{1}{2}\sum_k\Psi^{\dagger}_{k} \LBDG(k) \Psi_{k}  -\mathrm{iTr}M\mathbb{1},
\end{equation}
where we defined the momentum-space Nambu spinor
\begin{equation}
    \Psi_k := \begin{pmatrix}
        \hat{c}_{k, 1} \\
        \hat{c}_{k, 2} \\
        \hat{c}^{\dagger}_{-k,1} \\
        \hat{c}^{\dagger}_{-k,2}
    \end{pmatrix}
\end{equation}
and the $4\times 4$ BdG momentum-space Lindbladian
\begin{equation}\label{eq:mom_space_bdg_lindbladian}
    \LBDG(k) := \begin{pmatrix}
        Z(k) & 2Y(k) \\
        0 & -Z^T(-k)
    \end{pmatrix}.
\end{equation}
Each term here is a $2 \times 2$ matrix in the indices $\mu,\nu$ (and transposition of $Z(k)$ is just flipping $\mu,\nu$ indices).
Explicitly, we have
\begin{equation}\label{eq:mom_space_Z_matrix_definition}
    \begin{aligned}
        Z(k)_{\mu\nu} &= 4H(k)_{\mu\nu} -\mathrm{i} M(k)_{\mu\nu} - \mathrm{i}M(-k)_{\nu\mu} \\
        Y(k)_{\mu\nu} &= \mathrm{i}M(-k)_{\nu\mu} -\mathrm{i}M(k)_{\mu\nu}.
    \end{aligned}
\end{equation}
One must be careful in deriving these, as transposition of the real-space $M^T$ involves swapping \emph{both} $i, j$ and $\mu,\nu$. Also notice that $H^T(-k) = -H(k)$ but $Z^T(-k) \neq -Z(k)$.
\\ \\
\textbf{
\centering Diagonalization of the momentum-space BdG matrix}
\\

We now argue that diagonal components of $Z(k)$ are even in $k$. Assuming a quadratic (second-quantized) Hermitian Hamiltonian $H$ driving the coherent part of Lindblad dynamics [Eq.~\eqref{eq:lindblad_dynamics}], its terms will involve $\cdag_i c_j + \text{h.c.}$ (and, potentially, superconducting terms $\cdag_i\cdag_j+ \text{h.c.}$). Terms like $\gamma_{i,\mu}\gamma_{j,\mu}$ will not be present, as these cancel with the Hermitian conjugate term. Thus $H_{a, \mu\mu} = 0$ for any range index $a$ [see start of this section], and so $H(k)_{\mu\mu}=0$. With this, we have 
\begin{equation}
    Z(k)_{\mu\mu} = -M(k)_{\mu\mu} - M(-k)_{\mu\mu},
\end{equation}
which is even in $k$. Importantly, this means Eq.~\eqref{eq:mom_space_bdg_lindbladian} \emph{is traceless}. In fact one can see that
\begin{equation}
    \mathrm{det}(\LBDG(k)-\lambda\mathbb{1}) = \mathrm{det}(Z(k)-\lambda\mathbb{1)}\mathrm{det}(-Z(-k)-\lambda\mathbb{1})
\end{equation}
meaning eigenvalues come in pairs that are eigenvalues of $Z(k)$ and (opposites of eigenvalues of) $Z(-k)$. We now proceed by analogy with the diagonalization in real space. Define the matrix $S(k)$ such that 
\begin{equation}\label{eq:change_basis_rid_Y}
    S^{-1}(k)\LBDG(k)S(k) = \begin{pmatrix}
        Z(k) & 0 \\
        0 & -Z^T(-k)
    \end{pmatrix} := \tilde{\mathcal{L}}(k).
\end{equation}
We can surmise that this matrix takes the form 
\begin{equation}
    S(k) = \begin{pmatrix}
        1 & X(k)\\
        0 & 1
    \end{pmatrix} \implies Z(k)X(k) + X(k)Z^T(-k) = -2Y(k)
\end{equation}
in order for Eq.~\eqref{eq:change_basis_rid_Y} to hold. As for the real space case, this equation defines the matrix $X(k)$, which contains information about $Y(k)$.
We also define diagonalizing matrices $V(k)$ and $W(k)$ such that
\begin{equation}
    \begin{aligned}
        Z(k) = V^{-1}(k)\Lambda_z(k)V(k), \\
        \tilde{\mathcal{L}}(k) = W^{-1}(k)\Lambda(k)W(k),
    \end{aligned}
\end{equation}
where $\Lambda_z(k)$ is the diagonalized $Z(k)$, and
\begin{equation}
    \Lambda(k) = \begin{pmatrix}
        \Lambda_z(k) & 0 \\
        0 & -\Lambda_z(-k)
    \end{pmatrix}
\end{equation}
is diagonal and consists of pairs of eigenvalues of $Z(k)$ and (opposites of) $Z(-k)$, as discussed above.
With the above, $\LBDG$ is diagonalized by the combination of $S(k)$ and $W(k)$. This implies that
\begin{equation}
    \Lk = \Psi^\dagger_kS(k)W^{-1}(k)\Lambda(k)W(k)S^{-1}(k)\Psi_k  -\mathrm{iTr}M\mathbb{1}.
\end{equation}
Let 
\begin{equation}
    P(k) = W(k)S^{-1}(k) = \begin{pmatrix}
        V(k) & 0 \\
        0 & (V^{-1}(-k))^T
    \end{pmatrix}
    \cdot
    \begin{pmatrix}
        1 & -X(k) \\
        0 & 1
    \end{pmatrix} = \begin{pmatrix}
        V(k) & -V(k)X(k) \\
        0 & (V^{-1}(-k))^T
    \end{pmatrix}.
\end{equation}
Then we can define new Nambu spinors
\begin{equation}
    \begin{pmatrix}
        \hat{b}_{-k,1} \\
        \hat{b}_{-k,2} \\
        \hat{b}^\prime_{-k,1}\\
        \hat{b}^\prime_{-k,2}
    \end{pmatrix} = P(k)    \begin{pmatrix}
        \chat_{k,1} \\
        \chat_{k,2} \\
        \chatdag_{-k,1}\\
        \chatdag_{-k,2}
    \end{pmatrix}.
\end{equation}
[We choose this sign convention for the $\hat{b}$ operators to match the intuition of creation operators at a given $k$.] This yields (writing the $k$ variable as a subscript now to de-clutter)
\begin{equation}\label{eq:mom_space_b_operator_definitions}
    \begin{aligned}
        \hat{b}_{k, \mu} &= \sum_\nu \left[(V_{-k})_{\mu\nu}\chat_{-k,\nu} - (V_{-k}X_{-k})_{\mu\nu}\chatdag_{k,\nu}\right], \\
        \hat{b}^\prime_{k, \mu} &= \sum_\nu(V^{-1}_{k})^T_{\mu\nu}\chatdag_{k,\nu}.
    \end{aligned}
\end{equation}
Similarly to the real-space diagonalization, one can compute $P^{-1}$ acted on the row-vector of operators to obtain
\begin{equation}
    \Lk = \frac{1}{2}\begin{pmatrix}
        \underline{\hat{b}}^\prime_k, \quad \underline{\hat{b}}_k
    \end{pmatrix}
    \cdot \begin{pmatrix}
        \Lambda_z(k) & 0 \\
        0 & -\Lambda_z(-k)
    \end{pmatrix}
    \begin{pmatrix}
        \underline{\hat{b}}_{-k}\\
        \underline{\hat{b}}^\prime_{-k} 
    \end{pmatrix} - \mathrm{iTr}M\mathbb{1},
\end{equation}
 where the $\underline{\hat{b}}_k$ vectors are in $\mu = 1, 2$. As in real space, one can show that these operators satisfy almost-canonical commutation relations, now written as
 \begin{equation}
     \begin{aligned}
         \{\hat{b}_{k,\mu},\hat{b}_{q,\nu}\} &= 0 = \{\hat{b}^\prime_{k,\mu},\hat{b}^\prime_{q,\nu}\}, \\\{\hat{b}_{k,\mu},\hat{b}^\prime_{q,\nu}\} &= \delta_{-k, q}\delta_{\mu,\nu}.
     \end{aligned}
 \end{equation}
With these (and flipping the $k$ sum in the second term), we obtain
 \begin{equation}\label{eq:mom_space_diagonalized_bdg_lindbladian}
     \Lk = \sum_\mu\lambda_{k, \mu}\hat{b}^\prime_{k,\mu}\hat{b}_{-k,\mu},
 \end{equation}
 where $\lambda_{k,\mu}$ are the two eigenvalues of $Z(k)_{\mu\nu}$.
 Thus importantly, \emph{the eigenvalues of the BdG Lindbladian in momentum space are just the eigenvalues of the $2\times2$ matrix $Z(k)_{\mu\nu}$.}
 Excitations are created via the operators $\hat{b}^\prime_{k,\mu}$ on the non-equilibrium steady-state $\NESS$ and have eigenvalues $\lambda_{k,\mu}$.
\section{Lindbladian versus postselected non-Hermitian winding number}\label{sec:winding_derivation}
Our goal in this section is to obtain the eigenvalues $\lambda$ of the third-quantized $\Lplus$ in terms of $\Hpost$. To do so, we derive the action of the (non-vectorized) operator $\mathcal{L}$ on single superfermion states, and compare it to the third-quantized action of $\Lplus$. Firstly, we wish to find a simple way of extracting those eigenvalues. We should avoid any identities involving $\NESS$, as this is a complicated unknown state that involves arbitrarily many super-fermion operators. Instead, we use the fact that [see Eq.~\eqref{eq:mom_space_diagonalized_bdg_lindbladian}]
\begin{equation}\label{eq:extracting_lambda_eigenvalues}
\begin{aligned}
    \bbra{\mathbb{1}}\hat{b}_{k,\mu}\Lplus = \sum_{p,\nu} \lambda_{p,\nu}\bbra{\mathbb{1}}\hat{b}_{k,\mu}\hat{b}^\prime_{p,\nu}\hat{b}_{-p,\nu} = \bbra{\mathbb{1}}\hat{b}_{k,\mu}\lambda_{-k,\mu} \\ \implies
    \Lplus^\dagger \hat{b}^\dagger_{k,\mu}\kket{\mathbb{1}} = \lambda^*_{-k,\mu}\hat{b}^\dagger_{k,\mu}\kket{\mathbb{1}}.
\end{aligned}
\end{equation}
Thus, we may use the action of $\Lplus^\dagger$ on single-superfermion excitations to obtain the eigenvalues $\lambda_{k,\mu}$\footnote{There is a subtlety here: $\Lplus$ is the Lindbladian restricted to the positive super-fermion parity sector of $\mathcal{K}$, so how can it act on a single-particle excitation? To solve this, we should now regard Eq.~\eqref{eq:mom_space_diagonalized_bdg_lindbladian} as a definition for $\Lplus$ as a generic operator on $\mathcal{K}$.}.
We also use the decomposition (for an arbitrary $X \in \rho(\mathcal{H})$) 
\begin{equation}
\mathcal{L}X = \mathcal{L}_{\mathrm{post}}X + \mathcal{L}_{\mathrm{Q}}X,
\end{equation}
which in third quantization corresponds to the superoperator $\hat{\mathcal{L}}$. Here, $\mathcal{L}_{\mathrm{Q}}X := \sum_m J_m X \Jdag_m$ [see Eq.~\eqref{eq:schrodynamicsplusquantumjumps}]. When mapping these to super-operators, $\mathcal{L}_\mathrm{Q}$ is the only term that involves super-fermion parity [see Eq.s~\eqref{eq:first_step_deriving_jump_lindbladian} and~\eqref{eq:liouville_basis_identities}]. Indeed, using these equations, we can write the corresponding super-operator term as $\hat{\mathcal{L}}_{\mathrm{Q}} = e^{\mathrm{i}\pi\hat{\mathcal{N}}}\hat{Q}$. We then define the \emph{new} auxiliary super-operator 
\begin{equation}
    \Ltildehat := \hat{\mathcal{L}}_{\mathrm{post}} - e^{\mathrm{i}\pi\hat{\mathcal{N}}}\hat{Q}
\end{equation}
(i.e. negating the term involving super-parity compared to $\hat{\mathcal{L}}$). This is the super-operator counterpart of 
\begin{equation}
    \Ltilde X = \mathcal{L}_{\mathrm{post}} X - \LQ X.
\end{equation}
Then, using $e^{\mathrm{i}\pi\hat{\mathcal{N}}}\chat^{(\dagger)}_{i,\mu}e^{-\mathrm{i}\pi\hat{\mathcal{N}}} = -\chat^{(\dagger)}_{i,\mu}$ (and the fact that $\kket{\mathbb{1}}$ is an even-super-parity state), we have
\begin{equation}
\begin{aligned}
    \Ltildehat \chatdag_{i,\mu}\kket{\mathbb{1}} = \hat{\mathcal{L}}_{\mathrm{post}}\chatdag_{i,\mu}\kket{\mathbb{1}} + \hat{Q}\chatdag_{i,\mu}\kket{\mathbb{1}} := \Lplus \chatdag_{i,\mu}\kket{\mathbb{1}}
\end{aligned}
\end{equation}
i.e. the action of this operator on odd-superfermion parity states (and in particular on single-superfermion operators) is the same as that of $\Lplus$.
Then in particular the above\footnote{Using the definition whereby the action of the super-operator outside the ket corresponds to that of the regular operator inside the ket, i.e. $\Ltildehat \kket{\gamma_{i,\mu}} := \kket{\Ltilde\gamma_{i,\mu}}$.} implies that
\begin{equation}
   \Lplus \chatdag_{i,\mu}\kket{\mathbb{1}} = \kket{\Ltilde \gamma_{i,\mu}}.
\end{equation}
Along the same vein, one may easily show that the same holds for daggered superoperators, i.e.
\begin{equation}\label{eq:Lplus_is_Ltilde}
    \Lplus^\dagger \chatdag_{i,\mu}\kket{\mathbb{1}} = \kket{\Ltilde^\dagger \gamma_{i,\mu}}.
\end{equation}
We will return to this identity later.

\subsection{Second-quantized Lindbladian identities}
We consider both gain and loss jump operators, defined as in the main text via
\begin{equation}
\begin{aligned}
    L_m &= \sum_m \mathcal{L}_{mi}c_i, \\
    G_m &= \sum_m \mathcal{G}_{mi}\cdag_i.
\end{aligned}
\end{equation}
Before we even use the super-operator formalism, one can show
\begin{equation}
\begin{aligned}
    -\mathrm{i}\mathcal{L}X &:= -\mathrm{i}\left[H, X\right] + \sum_m \left( L_m X \Ldag_m - \frac{1}{2}\{\Ldag_m L_m, X\}\right)  + \sum_m \left(G_m X \Gdag_m - \frac{1}{2}\{\Gdag_m G_m, X\}\right) \\
    &= \sum_{ij}\left[-\mathrm{i}\mathcal{H}_{ij}\left[\cdag_ic_j, X\right] + \sum_m \left(\mathcal{L}_{mi}\mathcal{L}^*_{mj}c_i X \cdag_j - \frac{1}{2}\mathcal{L}^*_{mi}\mathcal{L}_{mj}\{\cdag_ic_j, X\}\right) + \text{(gain term)}\right] \\
    &= \sum_{ij}\left[-\mathrm{i}\mathcal{H}_{ij}\left[\cdag_ic_j, X\right] + m^{(l)}_{ij}\left(c_j X \cdag_i - \frac{1}{2}\{\cdag_ic_j, X\}\right)+ m^{(g)}_{ij}\left(\cdag_j X c_i - \frac{1}{2}\{c_i\cdag_j, X\}\right)\right]
\end{aligned}
\end{equation}
where we defined $m^{(l)}_{ij} = \sum_m \mathcal{L}^*_{mi}\mathcal{L}_{mj}, \; m^{(g)}_{ij} = \sum_m \mathcal{G}^*_{mi}\mathcal{G}_{mj}$.
By comparison, we also have
\begin{equation}
    \Hpost = H - \frac{\mathrm{i}}{2}\sum_m \Ldag_m L_m - \frac{\mathrm{i}}{2}\sum_m \Gdag_m G_m = \sum_{ij}\left[ \mathcal{H}_{ij} - \frac{\mathrm{i}}{2}m^{(l)}_{ij} \mp \frac{\mathrm{i}}{2}m^{(g)}_{ji} \right] \cdag_i c_j - \frac{\mathrm{i}}{2} \left[ \sum_i m^{(g)}_{ii}\right]\mathbb{1}
\end{equation}
where in the last equality we used commutation/anti-commutation relations (for bosons/fermions respectively), and hence we obtain $-$ for bosons, $+$ for fermions.
We define $\mathcal{H}^{\mathrm{post}}_{ij}$ and $\mathcal{H}^{\mathrm{post}}(k)$ such that
\begin{equation}\label{eq:Heff_plus_identity_definition}
    \Hpost = \sum_{ij}\mathcal{H}^{\mathrm{post}}_{ij}\cdag_i c_j + c\mathbb{1} = \sum_k \mathcal{H}^{\mathrm{post}}(k) \cdag_k c_k + c\mathbb{1},
\end{equation}
for $c$ some constant. We now use the definition of the super-operator inner product for arbitrary operators $X,Y$ [see also Appendix of Ref.~\cite{gideon2023_anomalousl_relaxations_NH}]
\begin{equation}
\begin{aligned}
    \bbra{\mathcal{L}^\dagger Y}X\rangle\!\rangle &:= \frac{1}{2^N} \mathrm{Tr}[(\mathcal{L}^\dagger Y)^\dagger X], \\
    \bbra{Y}\mathcal{L}X\rangle\!\rangle &:= \frac{1}{2^N} \mathrm{Tr}[Y^\dagger \mathcal{L}X] = \frac{1}{2^N}\mathrm{Tr}[\mathcal{L}XY^\dagger]
\end{aligned}
\end{equation}
and cyclicity of the trace to obtain
\begin{equation}
    \mathrm{i}\mathcal{L}^\dagger Y = \sum_{ij}\left[\mathrm{i}\mathcal{H}_{ij}\left[\cdag_ic_j, Y\right] + m^{(l)}_{ij}\left(\cdag_i Y c_j - \frac{1}{2}\{\cdag_ic_j, Y\}\right) + m^{(g)}_{ij}\left(c_i Y \cdag_j - \frac{1}{2}\{c_i\cdag_j, Y\}\right)\right].
\end{equation}
[Note that Gideon et al.'s version~\cite{gideon2023_anomalousl_relaxations_NH} of this disagrees with ours by a Hermitian transpose.] Now using the definition of the auxiliary operator $\Ltilde$ from the previous section, we see that
\begin{equation}
    \mathrm{i}\Ltilde^\dagger Y = \sum_{ij}\left[\mathrm{i}\mathcal{H}_{ij}\left[\cdag_ic_j, Y\right] + m^{(l)}_{ij}\left(-\cdag_i Y c_j - \frac{1}{2}\{\cdag_ic_j, Y\}\right) + m^{(g)}_{ij}\left(-\cdag_i Y c_j - \frac{1}{2}\{\cdag_ic_j, Y\}\right)\right],
\end{equation}
i.e. the quantum jump term involving $m^{(l)}_{ij}$ and $m^{(g)}_{ij}$ has an extra minus. We now use the following identities:
\begin{equation}
\begin{aligned}
    \left[\cdag_ic_j, c_l\right] &= \begin{cases}
        \delta_{il}c_j \text{ for bosons} \\
        -\delta_{il}c_j \text{ for fermions};
    \end{cases}\\
    -\cdag_ic_lc_j - \frac{1}{2}\{\cdag_ic_j, c_l\} &= \begin{cases}
        -2\cdag_ic_lc_j - \frac{1}{2}\delta_{il}c_j \text{ for bosons} \\
        -\frac{1}{2}\delta_{il}c_j \text{ for fermions};
    \end{cases} \\
    -c_ic_l\cdag_j - \frac{1}{2}\{c_i\cdag_j, c_l\}& = \begin{cases}
        -2c_ic_l\cdag_j + \frac{1}{2}\delta_{jl}c_i \text{ for bosons} \\
        -\frac{1}{2}\delta_{jl}c_i \text{ for fermions};
    \end{cases}
\end{aligned}
\end{equation}
and similar identities for $\cdag_l$ instead of $c_l$. 
These identities together with the above definition of $\Ltilde$ give the following: \emph{for bosons},
\begin{equation}
\begin{aligned}
\mathrm{i}\mathcal{L}^\dagger c_i &= \mathrm{i}\sum_j \left[\mathcal{H}_{ij} + \frac{\mathrm{i}}{2}m^{(l)}_{ij}-\frac{\mathrm{i}}{2}m^{(g)}_{ji}\right]c_j,\\
\mathrm{i}\mathcal{L}^\dagger \cdag_i &= -\mathrm{i}\sum_j \cdag_j \left[ \mathcal{H}_{ji} - \frac{\mathrm{i}}{2}m^{(l)}_{ji} +\frac{\mathrm{i}}{2}m^{(g)}_{ij}\right],
\end{aligned}
\end{equation}
whereas \emph{for fermions},
\begin{equation}
\begin{aligned}
    \mathrm{i}\Ltilde^\dagger c_{i} &= -\mathrm{i}\sum_{j}\left[\mathcal{H}_{ij} -\frac{\mathrm{i}}{2}m^{(l)}_{ij}-\frac{\mathrm{i}}{2}m^{(g)}_{ji}\right]c_{j}, \\
    \mathrm{i}\Ltilde^\dagger \cdag_{i} &= \mathrm{i}\sum_{j}\cdag_{j} \left[\mathcal{H}_{ji} +\frac{\mathrm{i}}{2}m^{(l)}_{ji} + \frac{\mathrm{i}}{2}m^{(g)}_{ij}\right].
\end{aligned}
\end{equation}
Notice for bosons we only needed to use $\mathcal{L}^\dagger$; we did not need to use $\Ltilde$ (in fact using that for bosons does not give such a nice expression). 

Comparing the above coefficients with $\mathcal{H}^{\mathrm{post}}_{ij}$, we notice a difference in sign for the loss terms (for bosons) and for the gain terms (for fermions)~\cite{mcdonald_ness_of_NH_models}. We introduce new, \emph{effective} single-particle Hamiltonians $\mathcal{H}^{\mathrm{eff}}_{\mathrm{B}}$ and $\mathcal{H}^{\mathrm{eff}}_{\mathrm{F}}$ (where $\text{B,F}$ stand for bosons and fermions respectively), defined as
\begin{equation}\label{eq:effective_hamiltonians_definitions}
\begin{aligned}
    (\mathcal{H}^{\mathrm{eff}}_{\mathrm{B}})_{ij} := \mathcal{H}_{ij} + \frac{\mathrm{i}}{2}m^{(l)}_{ij} - \frac{\mathrm{i}}{2}m^{(g)}_{ji}, \\
    (\mathcal{H}^{\mathrm{eff}}_{\mathrm{F}})_{ij} := \mathcal{H}_{ij} - \frac{\mathrm{i}}{2}m^{(l)}_{ij} - \frac{\mathrm{i}}{2}m^{(g)}_{ji}.
\end{aligned}
\end{equation}
We define the full, many-body effective Hamiltonians as $H^{\mathrm{eff}}_{\mathrm{B,F}} = \sum_{ij}(\mathcal{H}^{\mathrm{eff}}_{\mathrm{B,F}})_{ij}\cdag_ic_j$. With these definitions, we have \emph{for bosons}
\begin{equation}
\begin{aligned}
    \mathcal{L}^\dagger c_i &= \sum_j (\mathcal{H}^{\mathrm{eff}}_{\mathrm{B}})_{ij}c_j, \\
    \mathcal{L}^\dagger \cdag_i &= -\sum_j \cdag_j (\mathcal{H}^{\mathrm{eff}}_{\mathrm{B}})_{ij}^*,
\end{aligned}
\end{equation}
and \emph{for fermions}
\begin{equation}
\begin{aligned}
    \Ltilde^\dagger c_i &= -\sum_j (\mathcal{H}^{\mathrm{eff}}_{\mathrm{F}})_{ij}c_j, \\
    \Ltilde^\dagger \cdag_i &= \sum_j \cdag_j (\mathcal{H}^{\mathrm{eff}}_{\mathrm{F}})_{ij}^*.
\end{aligned}
\end{equation}
Fourier-transforming the fermions, one can also show
\begin{equation}
\begin{cases}
    \mathcal{L}^\dagger c_k = \mathcal{H}^{\mathrm{eff}}_{\mathrm{B}}(k)c_k, \quad 
    \mathcal{L}^\dagger \cdag_k = -\cdag_k \mathcal{H}^{\mathrm{eff}}_{\mathrm{B}}(k)^* \quad \text{   for bosons}, \\  
    \Ltilde^\dagger c_{k} = -\mathcal{H}^{\mathrm{eff}}_{\mathrm{F}}(k)c_{k}, \quad
    \Ltilde^\dagger \cdag_{k} = \cdag_{k}\mathcal{H}^{\mathrm{eff}}_{\mathrm{F}}(k)^* \quad \text{ for fermions.}
\end{cases}
\end{equation}
Here we only treat the fermion case; the bosonic case easily follows\footnote{While in this work we focus on the fermion case, it would be interesting to fully investigate the topological correspondence for bosons~\cite{Wanjura2020, PhysRevLett.127.213601, 10.21468/SciPostPhys.15.4.173}.}. We drop the subscript $\mathrm{F}$ to de-clutter. We recall the definition of Majoranas as
\begin{equation}
    c_i = \frac{1}{2}(\gamma_{i,1} - \mathrm{i}\gamma_{i,2}) \implies \begin{cases}
        \gamma_{i,1} = c_i + \cdag_{i} \\
        \gamma_{i,2} = \mathrm{i}(c_i - \cdag_{i}).
    \end{cases}
\end{equation}
This allows us to define Fourier-transformed Majoranas
\begin{equation}
    \gamma_{k,\mu} = \frac{1}{\sqrt{N}}\sum_j \gamma_{j,\mu}e^{\mathrm{i}kj} \implies \begin{cases}
        \gamma_{k,1} = c_k + \cdag_{-k} \\
        \gamma_{k,2} = \mathrm{i}(c_k - \cdag_{-k}).
    \end{cases}
\end{equation}
We then obtain
\begin{equation}\label{eq:Ltilde_on_mom_space_majoranas}
\begin{aligned}
    \Ltilde^\dagger \gamma_{k,1} &= -[c_k \mathcal{H}^{\mathrm{eff}}(k) - \cdag_{-k}\mathcal{H}^{\mathrm{eff}}(-k)^*], \\
    \Ltilde^\dagger \gamma_{k,2} &= -\mathrm{i}[c_k \mathcal{H}^{\mathrm{eff}}(k)  + \cdag_{-k}\mathcal{H}^{\mathrm{eff}}(-k)^*].
\end{aligned}
\end{equation}
We will use these identities later.
\subsection{Derivation of Lindbladian eigenvalues (for fermions)}
We now have all the ingredients to obtain the eigenvalues $\lambda_{k,\mu}$.
We use Eq.~\eqref{eq:Lplus_is_Ltilde} and the fact that $\chatdag_{k,\nu}\kket{\mathbb{1}} = \kket{\gamma_{-k,\nu}}$ to see that [see also Eq.~\eqref{eq:mom_space_b_operator_definitions}]
\begin{equation}
    \Lplus^\dagger \hat{b}^\dagger_{k,\mu}\kket{\mathbb{1}} = \sum_{\nu}(V^*_{-k})_{\mu\nu}\Lplus^\dagger\chatdag_{-k,\nu}\kket{\mathbb{1}} = \sum_{\nu}(V^*_{-k})_{\mu\nu}\kket{\Ltilde^\dagger \gamma_{k,\nu}}.
\end{equation}
But also by Eq.~\eqref{eq:extracting_lambda_eigenvalues}, we have
\begin{equation}
    \lambda^*_{-k,\mu}\sum_{\nu}(V^*_{-k})_{\mu\nu}\kket{\gamma_{k,\nu}} =\sum_{\nu}(V^*_{-k})_{\mu\nu}\kket{\Ltilde^\dagger \gamma_{k,\nu}}.
\end{equation}
Using this and Eq.~\eqref{eq:Ltilde_on_mom_space_majoranas}, we obtain the matrix equation\footnote{We use the important corollary that if $A\kket{\gamma_{k,1}} + B\kket{\gamma_{k,2}} = C\kket{\gamma_{k,1}} + D\kket{\gamma_{k,2}}$, then plugging in Fourier transforms and using the fact that $\kket{\gamma_{i,\mu}}$ are part of the canonical basis~\eqref{eq:canonical_liouville_basis} of the space $\mathcal{K}$, we may simply read off components as $A=C, B=D$.}
\begin{equation}
    \frac{1}{2}\begin{pmatrix}
        -(\mathcal{H}^{\mathrm{eff}}(k) - \mathcal{H}^{\mathrm{eff}}(-k)^*)& -\im(\mathcal{H}^{\mathrm{eff}}(k) + \mathcal{H}^{\mathrm{eff}}(-k)^*) \\
        \im(\mathcal{H}^{\mathrm{eff}}(k) + \mathcal{H}^{\mathrm{eff}}(-k)^*) & -(\mathcal{H}^{\mathrm{eff}}(k) - \mathcal{H}^{\mathrm{eff}}(-k)^*)
    \end{pmatrix}\begin{pmatrix}
        (V^*_{-k})_{\mu1} \\
        (V^*_{-k})_{\mu2}
    \end{pmatrix} = \lambda^*_{-k,\mu}\begin{pmatrix}
        (V^*_{-k})_{\mu1} \\
        (V^*_{-k})_{\mu2}        
    \end{pmatrix}.
\end{equation}
We simply need to diagonalize this matrix (as well as conjugate and let $-k \rightarrow k$) to obtain
\begin{equation}\label{eq:Z_eigenvalues_are_H_eff_eigenvalues}
    \boxed{
    \lambda_{k,1} = \mathcal{H}^{\mathrm{eff}}(k), \quad \lambda_{k,2} = -\mathcal{H}^{\mathrm{eff}}(-k)^*.
    }
\end{equation}
Thus an important fact: \emph{the eigenvalues of $\Lplus$ are just the eigenvalues $\lambda_{k,\mu}$ of the $Z(k)_{\mu\nu}$ matrix in Eq.~\eqref{eq:mom_space_bdg_lindbladian}, which are themselves entirely determined by the single-particle momentum-space effective/unconditional Hamiltonian} $\mathcal{H}^{\mathrm{eff}}(k)$. Eq. (9) in the main text is another way of stating this result for open boundary conditions in real space, by using a combined label $\alpha = 1, \ldots 2N$ for the label $(i, \mu)$. Notice the effective Hamiltonian only coincides with the conditional Hamiltonian in the absence of gain, i.e. $\Heff = \Hpost$ only if $G_m = 0$.
If we redefine our Hermitian part of the Hamiltonian $H \rightarrow H + c\mathbb{1}$, this leaves everything unchanged, even the Lindbladian super-operators, as we only ever compute commutators $\left[H, X\right]$. Yet, the above result still holds due to the extra $c\mathbb{1}$ term in Eq.~\eqref{eq:Heff_plus_identity_definition}. So we must be careful: we do not simply have the freedom to shift the Lindbladian spectrum by any constant; the $Z(k)$ spectrum will remain unchanged. It only cares about the “nontrivial" terms in the effective Hamiltonian.
\subsection{Winding numbers}
Let us define winding numbers of a momentum-space single-particle Hamiltonian $\mathcal{H}(k)$ and of $Z(k)$ as
\begin{equation}
\begin{aligned}
    \nu(E) &= \frac{1}{2\pi\mathrm{i}}\int_{-\pi}^\pi \mathrm{d}k \partial_k\log\det(\mathcal{H}(k) - E), \\
    \nu_Z(E) &= \frac{1}{2\pi\mathrm{i}}\int_{-\pi}^\pi \mathrm{d}k \partial_k\log\det(Z(k) - E\mathbb{1}),
\end{aligned}
\end{equation}
where as in the main text $E\in \mathbb{C}$ is a complex energy. Then we use
\begin{equation}
    \det(Z(k)-E\mathbb{1}) = \det \begin{pmatrix}
        \mathcal{H}^{\mathrm{eff}}(k) - E & 0 \\
        0 & -\mathcal{H}^{\mathrm{eff}}(-k)^* - E
    \end{pmatrix} = \det(\mathcal{H}^{\mathrm{eff}}(k) - E)\det(-\mathcal{H}^{\mathrm{eff}}(-k)^* - E).
\end{equation}
Using the fact that the logarithm of a product is the sum of logarithms, the winding number $\nu_Z(E)$ splits into
\begin{equation}
\begin{aligned}
    \nu_Z(E) &= \frac{1}{2\pi\mathrm{i}}\int_{-\pi}^\pi \mathrm{d}k \partial_k[\log\det(\mathcal{H}^{\mathrm{eff}}(k) - E) +\log\det(-\mathcal{H}^{\mathrm{eff}}(-k)^* - E)]
\end{aligned}
\end{equation}
The first term is just $\nueff(E)$, the winding number of $\mathcal{H}^{\mathrm{eff}}$; the second is more subtle. Letting $k \rightarrow -k$, the integral bounds flip (thus negating the whole thing); also we pull out a $\log \mathrm{i}^2$ which annihilates with the derivative $\partial_k$. Thus the second term is
\begin{equation}
    -\frac{1}{2\pi\mathrm{i}}\int_{-\pi}^\pi \mathrm{d}k [\partial_k\log\det(\mathcal{H}^{\mathrm{eff}}(k) + E^*)]^* := \nueff(-E^*)^*.
\end{equation}
Using the fact that these winding numbers are real (and so $\nueff(-E^*)^* = \nueff(-E^*)$), we thus arrive at
\begin{equation}\label{eq:winding_relations}
    \boxed{
    \nu_Z(E) = \nueff(E) + \nueff(-E^*).
    }
\end{equation}
This equation is the root of all the freedom we have in engineering models with skin effects surviving/reversing/cancelling in the Lindbladian vs. $\Hpost$ picture. We did not refer to $\nu_Z$ in the main text because the block-diagonalization of $Z$ is completely $k$-independent and always possible in the presence of $U(1)$ symmetry, so even when $\nu_Z = 0$, a $\Heff$ skin effect will be stable and protected.
\subsection{Bulk-boundary correspondence in semi-infinite systems}
We now discuss consequences in semi-infinite systems. Importantly, in the third-quantization formalism, the Lindbladian super-operator is nothing but a non-Hermitian matrix [see Eq.~\eqref{eq:mom_space_bdg_lindbladian}]; thus using results on Toeplitz matrices from Ref.~\cite{Reichel_1992} (and already used for non-Hermitian Hamiltonians in Ref.~\cite{Okuma_2020}), we can immediately establish a bulk-boundary correspondence for the Lindbladian: in semi-infinite boundary conditions (SIBCs), the single-particle spectrum consists of the PBC single-particle spectrum together with its whole enclosing area in the complex plane, where the winding number is non-zero. This means that we need only look at the loops in the complex plane formed by the PBC spectrum $\{\lambda_{k,\mu}\}$ of $Z$ to determine allowed decaying edge states. In other words, the $Z$ matrix in PBC/SIBC is a circulant/Toeplitz matrix. We can use all the results from the literature on Toeplitz matrices~\cite{toeplitz_matrices_bottcher} to determine index theorems in our case~\cite{Okuma_2020}.
\begin{figure}
    \centering
    \includegraphics[scale=0.19]{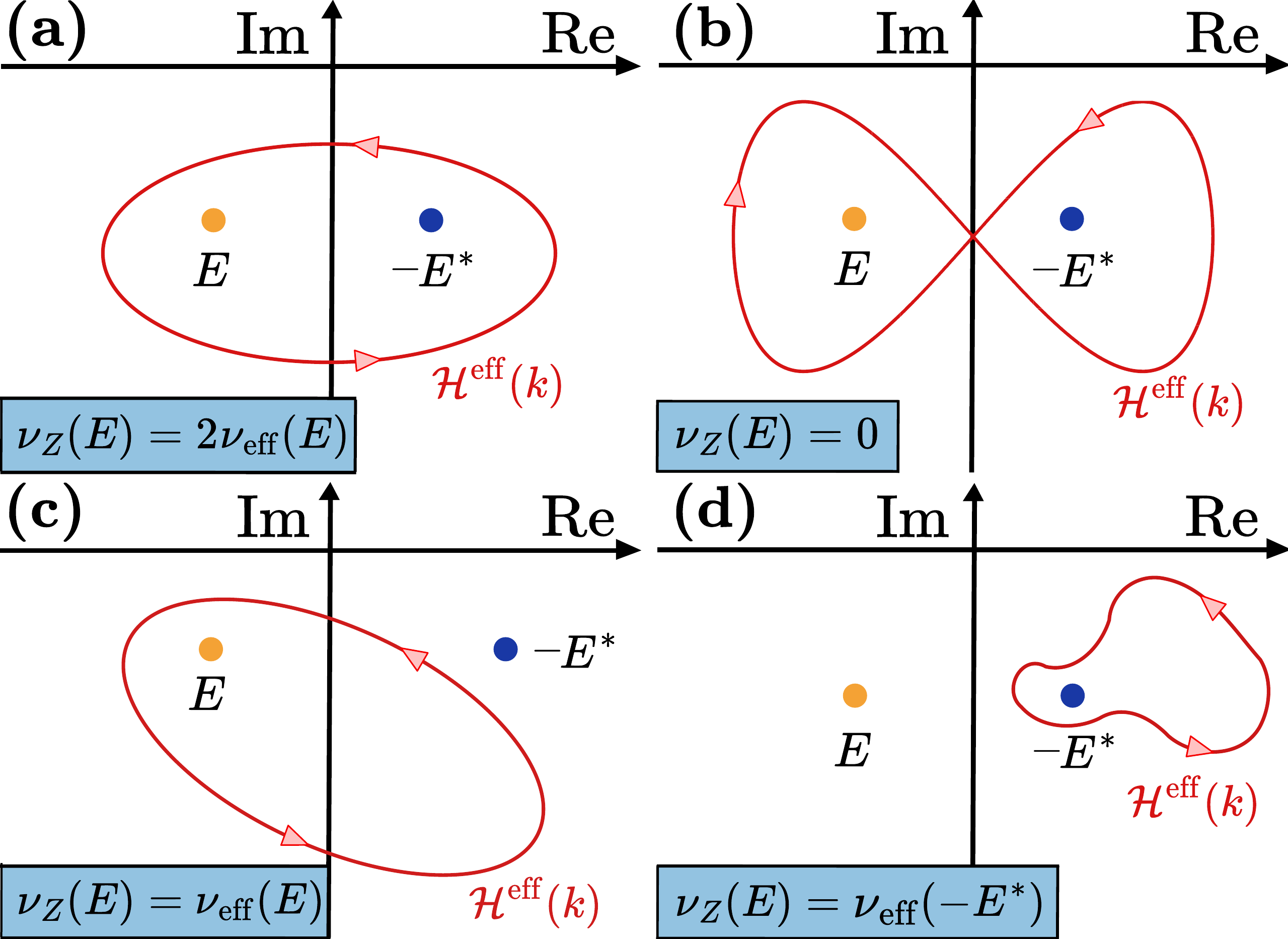}
    \caption{All possible cases for the winding number of the Lindbladian compared to the winding number of the effective/unconditional Hamiltonian $\mathcal{H}^{\mathrm{eff}}$. \textbf{(a)}: when the spectrum is a simple loop that is symmetric about the imaginary axis, the winding number is always double (e.g. the Hatano-Nelson model falls here); \textbf{(b)}: one can also have a self-intersecting loop which results in opposite windings; \textbf{(c)}: introducing imaginary coefficients in the Hamiltonian can “tilt" the ellipse, resulting in a winding number that can be equal to that of the Hamiltonian; \textbf{(d)}: introducing a constant on-site energy can shift the loop along the real axis, which is another way of introducing an asymmetry. Cases \textbf{(c)} and \textbf{(d)} both break $\text{PHS}^\dagger$ symmetry.}
    \label{fig:all_winding_cases}
\end{figure}
This has important physical consequences, as illustrated in Fig.~\ref{fig:all_winding_cases}. Reflection symmetry about the imaginary axis corresponds to the $\text{PHS}^\dagger$ symmetry in the 38-fold way classification of non-Hermitian Hamiltonians~\cite{kawabata_38_fold_way}, namely
\begin{equation}
P\mathcal{H}^{\mathrm{eff}}(k)^*P^{-1} = -\mathcal{H}^{\mathrm{eff}}(-k). 
\end{equation}
[Note e.g. for the Hatano-Nelson model, we must shift momenta by $\pi/2$ for this to work.]
If we also require no self-intersections, such symmetric spectra correspond to Fig~\ref{fig:all_winding_cases}(a). This case guarantees that the two windings on the right-hand-side of Eq.~\eqref{eq:winding_relations} are equal in magnitude.
Then whenever $\nueff(E) \neq 0$, the Lindblad winding $\nu_Z(E) = 2\nueff(E)$.
If we relax the restriction that there be no self-intersections, we can have a 'figure-eight' type case [see Fig.~\ref{fig:all_winding_cases}(b)] where the two windings cancel, and $\nu_Z(E) = 1 - 1 = 0$.
If we now relax the $\text{PHS}^\dagger$ symmetry, the two windings on the right-hand-side of Eq.~\eqref{eq:winding_relations} can be different in magnitude. The simplest example of this occurs by adding an on-site energy to the model, which shifts the spectrum along the real axis, as in Fig~\ref{fig:all_winding_cases}(d). Then, one can have $\nu_Z(E) = \nueff(E)$ or $\nu_Z(E) = \nueff(-E^*)$. The latter case shows that it is not sufficient to consider the value of $E$ \emph{within} the $\mathcal{H}^{\mathrm{eff}}(k)$ loop for non-trivial windings; we must also consider those within its particle-hole-symmetric region. 

As discussed above, a direct analogy between Hamiltonians and Lindbladians carries through. However we must remember that single-“particle" excitations of the Lindbladians are unphysical as they have a single super-fermion. Physical states will be made of an even number of super-fermions. However, if we do have a bulk-boundary correspondence for single-super-fermion modes, then constructing a state with two such edge modes will itself be an edge mode, so the discussion carries through.

With the aforementioned index theorems, our discussion shows that localized eigenstates in SIBCs can have quite different properties in the postselected versus full Lindbladian picture.
We leave further investigation of the comparisons for postselected versus full Lindbladian (as well as just effective Hamiltonian versus full Lindbladian) skin effects in SIBC to future work.

\section{Explicit example: Lindbladian `Hatano-Nelson' model}\label{sec:explicit_example}
We now discuss the explicit model introduced in the main text in further detail.

\subsection{Effective Hamiltonian}
We provide a detailed analysis of the simple model in the main text. To reiterate, we consider the simplest model that exhibits the non-Hermitian skin effect: the Hatano-Nelson model
\begin{equation}\label{eq:hatano_nelson}
\Hpost = \sum_{i=1}^N \left[ (t+g)\cdag_{i+1} c_i + (t-g) \cdag_i c_{i+1} \right],
\end{equation}
which is non-Hermitian via the $g$ terms. We wish to construct an open quantum system, i.e. to find a Hermitian Hamiltonian $H$ and jump operators, such that this system's conditional Hamiltonian is $\Hpost$. The obvious choice for the Hermitian part is 
\begin{equation}
H = \sum_i t \left[\cdag_{i+1} c_i + \cdag_i c_{i+1} \right].
\end{equation}
As in the main text, we now define loss and gain operators on each fermionic site as 
\begin{equation}\label{eq:simple_model_jump_operators}
\begin{aligned}
    L_i &= \sqrt{\gamma_l}(c_i - \mathrm{i}c_{i+1}), \\
    G_i &= \sqrt{\gamma_g}(\cdag_i - \mathrm{i}\cdag_{i+1}).
\end{aligned}
\end{equation}
This yields loss and gain matrices, $L = \sum_i\Ldag_i L_i, \; G=\sum_i \Gdag_i G_i$, which become
\begin{equation}
\begin{aligned}
    L &= 2\gamma_l\hat{N}+\mathrm{i}\gamma_l\sum_{i=1}^N(\cdag_{i+1}c_i -\cdag_ic_{i+1}), \\
    G &= 2\gamma_gN\mathbb{1}-2\gamma_g\hat{N}+\mathrm{i}\gamma_g\sum_{i=1}^N(\cdag_{i+1}c_i -\cdag_ic_{i+1}),
\end{aligned}
\end{equation}
where $\hat{N}=\sum_i\cdag_i c_i$ is the number operator\footnote{We put a hat on this operator to distinguish it from the other term in $G$ proportional to $N\mathbb{1}$.}. We now obtain the conditional Hamiltonian
\begin{equation} \label{eq: hpost new try}
\begin{aligned}
    \Hpost &= H - \frac{\mathrm{i}}{2}(L+G)\\
    &= \sum_i(t+\frac{\gamma_l + \gamma_g}{2})\cdag_{i+1}c_i + (t-\frac{\gamma_l+\gamma_g}{2})\cdag_ic_{i+1} -\mathrm{i}(\gamma_l-\gamma_g)\hat{N} - \mathrm{i}\gamma_gN\mathbb1.
\end{aligned}
\end{equation}
If we require that $\gamma_l + \gamma_g = 2g$, this does indeed reproduce our target Hatano-Nelson model, up to some extra terms that do not alter the presence of a skin effect. By contrast the effective Hamiltonian is [see Eq.~\eqref{eq:effective_hamiltonians_definitions} and discussion surrounding it]
\begin{equation}
\Heff = \sum_i(t+\frac{\gamma_l - \gamma_g}{2})\cdag_{i+1}c_i + (t-\frac{\gamma_l-\gamma_g}{2})\cdag_ic_{i+1} -\mathrm{i}(\gamma_l+\gamma_g)\hat{N}.
\end{equation}
Importantly, the effective Hamiltonian does \emph{not} have the extra extensive term $-\mathrm{i}\gamma_g N \mathbb1$ that (minus) $\Hpost$ has, as one would obtain by naïvely computing $\Heff$ as $H-\frac{\mathrm{i}}{2}(L-G)$. This guarantees that the single-particle spectrum of $\Heff$ remain negative in imaginary component, as required to avoid amplification of modes.

We note in passing that Eq.~\eqref{eq:effective_hamiltonians_definitions} together with the definition $H^{\mathrm{eff}}_{\mathrm{B,F}} = \sum_{ij}(\mathcal{H}^{\mathrm{eff}}_{\mathrm{B,F}})_{ij}\cdag_ic_j$ implies that the definitions
\begin{equation}
\begin{aligned}
H^{\mathrm{eff}}_{\mathrm{F}} = H-\frac{\mathrm{i}}{2}(L-G), \\
H^{\mathrm{eff}}_{\mathrm{B}} = H+\frac{\mathrm{i}}{2}(L+G),
\end{aligned}
\end{equation}
actually hold in general for quadratic Lindblad systems, not only in the present model.

Using the above, one can Fourier transform and easily retrieve the expressions in the main text (reiterated here for convenience):
\begin{equation}\label{eq:simple_gainloss_model_hamiltonians}
\begin{aligned}
  \mathcal{H}^{\mathrm{post}}(k) &= 2t\cos k + \mathrm{i}\gamma\sin k -\mathrm{i}\delta, \\
  \mathcal{H}^{\mathrm{eff}}(k) &= 2t\cos k + \mathrm{i}\delta\sin k -\mathrm{i}\gamma,
\end{aligned}
\end{equation}
where $\gamma:=\gamma_l+\gamma_g, \; \delta := \gamma_l-\gamma_g$. Then, as discussed in the main text, for $\gamma_l > \gamma_g$, both the conditional and effective Hamiltonians wind the same way; however for $\gamma_g > \gamma_l$, they wind oppositely, implying a Lindbladian skin reversal. There is a transition when $\gamma_g = \gamma_l$, when the point-gap of $\Heff$ closes, while that of $\Hpost$ remains open, implying a Lindbladian skin effect cancellation.

\begin{figure}
    \centering
    \includegraphics[width=0.95\linewidth]{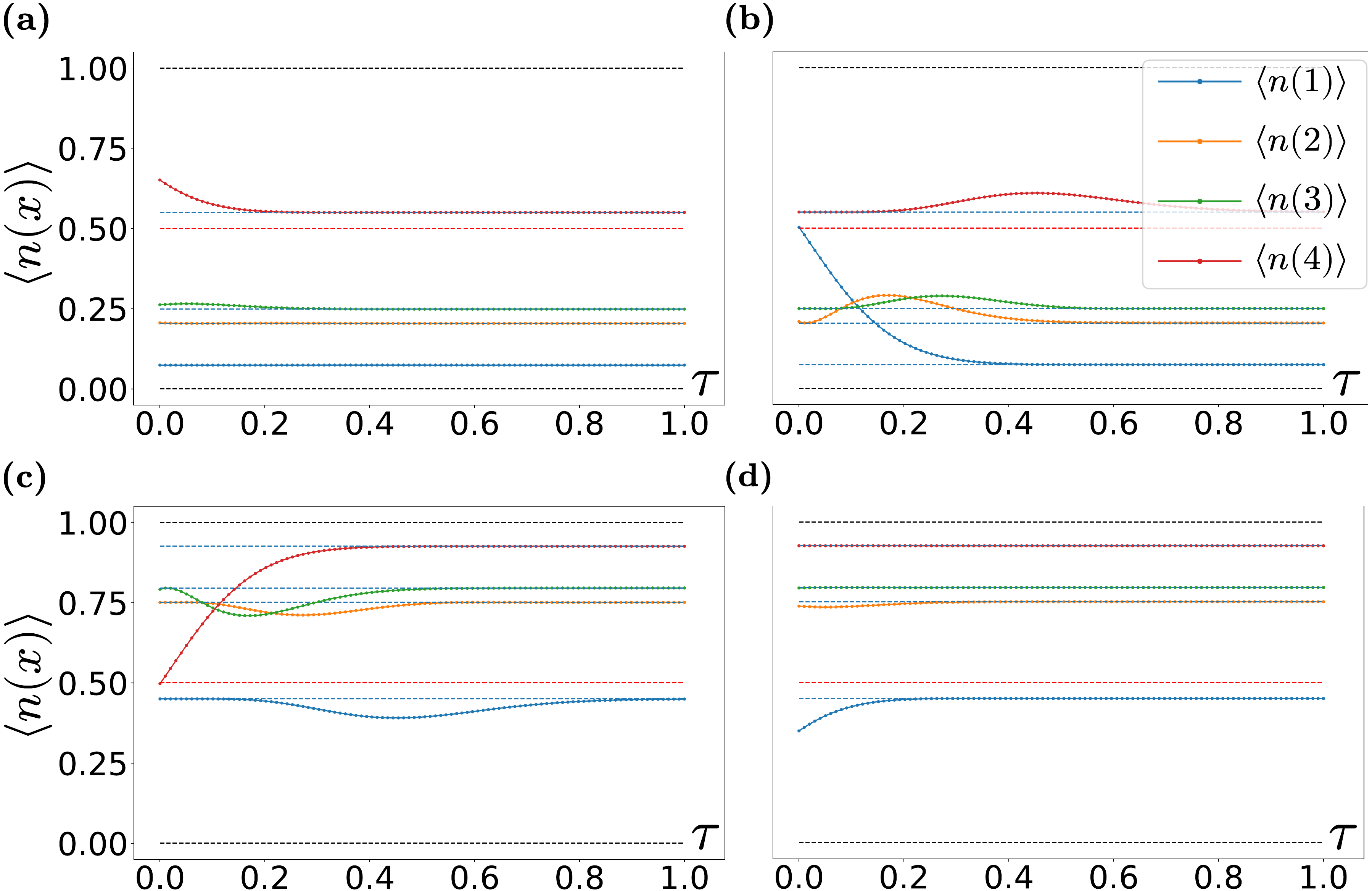}
    \caption{Plots of time (labelled as $\tau$ to avoid confusion with the hopping parameter $t$) evolution of a perturbation from the NESS, localized either on the left/right edge of the system, for $N=4$ fermions, $t=5$, and $\epsilon=0.5$ in Eq.~\eqref{eq:explicit_perturbation_form}. $\braket{n(i)}$ denotes the density on site $i = 1 \dots 4$ of the chain, where $i=1$ and $i=4$ corresponds to the left and right edge, respectively. (a) Evolution for $\gamma_l = 5, \gamma_g=1$, for a perturbation localized on the right edge. We observe simple exponential decay, and a small back-propagation due to Hermitian hopping in $t$. (b) Evolution for $\gamma_l=5, \gamma_g = 1$, for a perturbation localized on the left edge. There is a long-lived, right-moving perturbation, matching the fact that $\nu_{\mathrm{eff}}=1$. (c) Evolution for $\gamma_l = 1, \gamma_g = 5$, with a perturbation localized on the right edge. There is a long-lived, left-moving perturbation, matching the fact that now $\nu_{\mathrm{eff}}=-1$. (d) Evolution for $\gamma_l=1, \gamma_g=5$, for a perturbation localized on the left edge. There is mostly exponential decay here too. Dashed blue lines represent each site's density in the NESS; the dashed black lines represent the limits of minimal density $0$ and maximal density $1$, and the dashed bright red line represents half-filling.}
    \label{fig:explicit_model_dynamics}
\end{figure}

In Fig.~\ref{fig:explicit_model_dynamics}, we numerically compute (i.e. explicitly solve the differential equation for) Lindbladian dynamics for various localized perturbations away from the NESS, both for the loss-dominated and gain-dominated régime. For the loss-dominant case, our chosen perturbation takes the form (time is labelled by $\tau$)
\begin{equation}\label{eq:explicit_perturbation_form}
\rho(\tau=0) = \rho_{\mathrm{NESS}} + \epsilon \cdag_j \rho_{\mathrm{NESS}} c_j - \epsilon (1-\langle n_j \rangle_{\mathrm{NESS}})\rho_{\mathrm{NESS}}
\end{equation}
where the site index $j$ is $1$ ($N$) for a left- (right-) localized perturbation, respectively. This form ensures the initial density matrix (1) has unit trace, (2) is Hermitian, and (3) is positive semi-definite for small enough values of the real parameter $\epsilon$. For the gain-dominated case, we create hole-like excitations by replacing $\cdag_j$ with $c_j$ and vice-versa in the above formula. This choice for perturbations stems from our expectation that excitations are particle-like for $\gamma_l > \gamma_g$, and hole-like for $\gamma_g > \gamma_l$. Other choices do not affect our interpretation. Explicit dynamics show that an excitation localized on the ``wrong" side of the system (left for $\gamma_l > \gamma_g$, right for $\gamma_g > \gamma_l$) will travel to the opposite side of the system while decaying. This chiral motion would not occur if the perturbation were an eigenmode. On the other hand, an excitation on the ``correct" side (right for $\gamma_l > \gamma_g$, left for $\gamma_g > \gamma_l$) does exhibit simple exponential decay back to the NESS. This is expected for true eigenmodes of the Lindbladian, and matches the winding number discussion above. 

\begin{figure}
    \centering
    \includegraphics[width=0.95\linewidth]{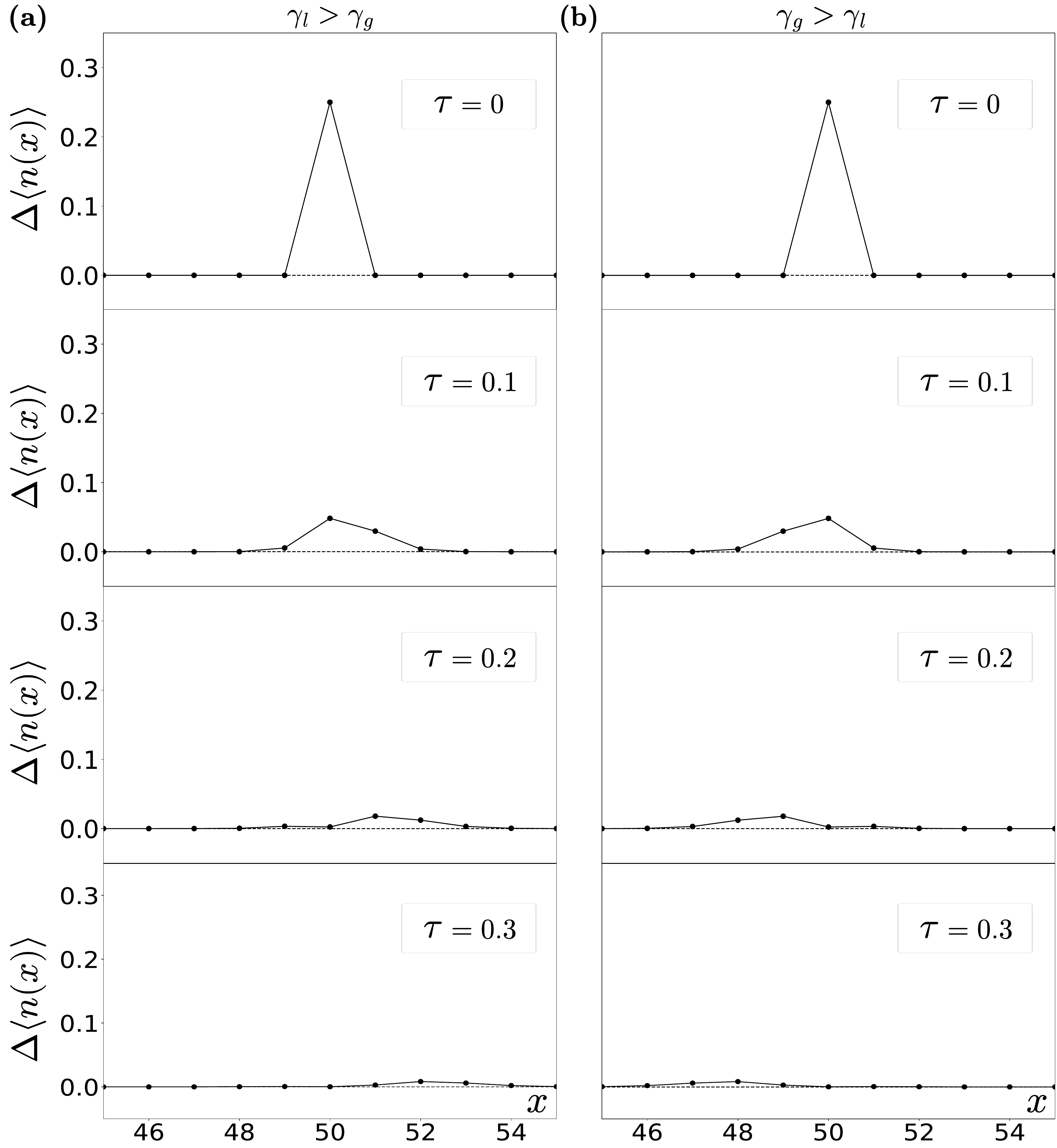}
    \caption{Time evolution of the difference in expectation of density to the NESS $\Delta\langle n(x) \rangle := \langle n(x) \rangle - \langle n(x)\rangle_{\mathrm{NESS}}$, for an initial perturbed density matrix with bump in the middle of the chain. We consider a chain of $N=100$ fermions, with an initial perturbation $\Delta \langle n(x=50)\rangle = 1/4$, and tight-binding hopping parameter $t=5$. We restrict attention to the vicinity of the middle of the chain $45 \leq x \leq 55$, where non-trivial evolution takes place given this initial condition. The horizontal black dashed line corresponds to $\Delta \langle n(x) \rangle = 0$, i.e., no deviation from the NESS density. (a) (left-hand vertical panels) Evolution for $\gamma_l = 5, \gamma_g=1$, corresponding to a winding number $\nu_\mathrm{eff} = +1$. There is clear rightward propagation, accompanied by a decay of the density towards the NESS density. (b) (right-hand vertical panels) Evolution for $\gamma_l=1, \gamma_g = 5$, corresponding to a winding number $\nu_\mathrm{eff} = -1$. There is clear leftward propagation, accompanied by a decay of the density towards the NESS density. We note that creating a density \emph{trough} compared to the NESS (e.g. $\Delta \langle n(x=50)\rangle < 0$) results in an analogous behavior; the direction of propagation is unchanged.}
    \label{fig:perturbation_animation}
\end{figure}

In Fig.~\ref{fig:perturbation_animation}, we also plot the density of perturbations as a function of position for different times $T$. As a sanity check of Fig.~\ref{fig:explicit_model_dynamics}, we perform this not by brute-force integration but by numerically solving the Lyapunov equation for the correlation matrix $\Delta_{ij} = \langle \cdag_i c_j \rangle$, which as a matrix equation is given by~\cite{mcdonald_ness_of_NH_models, PhysRevLett.123.170401} [see also Eq. (10) in the main text]
\begin{equation}
    \frac{\mathrm{d\Delta}}{\mathrm{d}\tau} = \im(\mathcal{H}^{\mathrm{eff}})^*\cdot \Delta - \im\Delta \cdot (\mathcal{H}^{\mathrm{eff}})^T + m^{(g)}.
\end{equation}
The NESS correlation matrix $\Delta_{\mathrm{NESS}}$ is obtained by setting the left-hand side of this equation to zero. We consider a simple perturbation by adding a “bump" in the middle of the chain, i.e. initializing the correlation matrix as that of the NESS, plus a perturbation in the diagonal element at the middle of the chain [see caption of Fig.~\ref{fig:perturbation_animation}]. We then see the direction in which this bump propagates, while also decaying in amplitude. Our findings corroborate our discussion above: the loss-dominant case sees rightward propagation, while the gain-dominant case sees leftward propagation, as implied by their non-zero winding number $\pm 1$.

Fig.~\ref{fig:perturbation_animation} illustrates the surprising fact that when $\gamma_g > \gamma_l$, a perturbation that increases electron density travels to the left, which is the opposite of what a naive interpretation of post-selected dynamics would suggest. This can be seen by the following physical argument.

Regardless of gain and loss parameters, one can perturb the NESS $\rho_{\mathrm{NESS}}$ to add localized electron density as in Eq.~\eqref{eq:explicit_perturbation_form}. Why does this bump travel to the left edge for $\gamma_g > \gamma_l$? The explanation boils down to the fact that when gain dominates, \emph{creating localized density above the NESS corresponds to removing some density of holes in the classical mixture making up the density matrix}. In the pure-state/post-selected picture, these holes then move to the left. Below we work out this picture mathematically.

In the gain-dominated case, the NESS is a mixture mostly of high-particle-number sectors.
For instance, when $\gamma_g \gg \gamma_l$, we may write to leading order
\begin{equation}
    \rho_\mathrm{NESS} = a^\mathrm{NESS}\rho_N + \sum_{j,k=1}^N b^\mathrm{NESS}_{jk}c_j\rho_N\cdag_k + \dots,
\end{equation}
where $a^\mathrm{NESS}$ and $b^\mathrm{NESS}_{jk}$ are scalars, and $\rho_N$ is the full-state density matrix $\rho_N = \ket{F}\bra{F}$ where $\ket{F} = \cdag_1\cdag_2\ldots\cdag_N \ket{0}$ is the fully occupied state. The perturbation 
\begin{equation}\label{eq:density_increasing_perturbation}
    \rho = (1-\epsilon + \epsilon\langle n_i\rangle_\mathrm{NESS})\rho_\mathrm{NESS} + \epsilon \cdag_i \rho_\mathrm{NESS} c_i.
\end{equation}
has a higher average particle number than the NESS, with a higher contribution to the full-particle-number sector, and lower contribution from the $N-1$ particle number sector (verified numerically). Let us consider evolution of this initial perturbation by writing to leading order
\begin{equation}\label{eq:time_evolution_of_perturbation}
    \rho(\tau) = a(\tau)\rho_N + \sum_{jk} b_{jk}(\tau)c_j\rho_N\cdag_k,
\end{equation}
with the initial density matrix being Eq.~\eqref{eq:density_increasing_perturbation}, i.e. $\rho(\tau=0) = \rho$.
In the full-particle sector (i.e. for $\ket{F}$), post-selected dynamics are trivial. The stochastic average (i.e. the full Lindblad evolution) will entail a simple relaxation from $a(0)\rho_N$ to $a^\mathrm{NESS}\rho_N$. 

On the other hand, in the sector of $(N-1)$ particles, the initial perturbation corresponds to a localised dip in the diagonal elements $b_{ii}$ compared to the NESS. As this is a dip in the probability for a hole at position $i$, then $\rho$ overall has increased density at position $i$ compared to the NESS. The full Lindblad dynamics can be unraveled into a postselected part within each sector of fixed particle number, and a jump part that connects the different sectors. In the present case, the jump part simply reduces $a$ and increases $b$ in Eq.~\eqref{eq:time_evolution_of_perturbation} back to their NESS values. Additionally, we next show that the postselected part moves an initial dip in $b_{ii}$ to the left. Consider a general single-hole state $\ket{\psi(\tau)}=\sum_{j=1}^N \beta_j(\tau)c_j\ket{F}$, with complex time-dependent coefficients $\beta_j(\tau)$. Comparing with Eq.~\eqref{eq: hpost new try}, the single-particle post-selected Hamiltonian matrix $\mathcal{H}^{\mathrm{post}}_{nm}$ has components $\mathcal{H}^{\mathrm{post}}_{m, m+1}=t-g$ and $\mathcal{H}^{\mathrm{post}}_{m+1, m}=t+g$ for all integers $m=1,\ldots N$. Under post-selected dynamics, $\ket
\psi$ satisfies
\begin{equation} \label{eq: postselbeta}
\begin{aligned}
    \partial_\tau\ket{\psi} &= -\im \Hpost \ket{\psi} \\ &= -\im\sum_{j,n,m}\mathcal{H}^{\mathrm{post}}_{nm}\cdag_n c_m \beta_j c_j \ket{F} \\
    &= +\im\sum_{j,m}\mathcal{H}^{\mathrm{post}}_{jm}\beta_jc_m\ket{F},
\end{aligned}
\end{equation}
where we have used that $\mathcal{H}^{\mathrm{post}}$ does not contain diagonal elements in the second line.
In other words, $\partial_\tau \bs{\beta} = +\im \bs{\beta}^{\mathrm{T}} \cdot  \mathcal{H}^{\mathrm{post}}$ as a row-vector; explicitly here this gives $\partial_\tau\beta_j = \im\left[\beta_{j-1}(t-g)+\beta_{j+1}(t+g)\right]$, i.e. hopping is stronger to the left. In particular, note that this implies that when we start from a $\bs{\beta}$ profile that is homogenous except for a localised dip somewhere, then by the linearity of the evolution equation \emph{this dip will also move to the left} (verified numerically). This result is non-intuitive: the second-quantised Hamiltonian Eq.~\eqref{eq: hpost new try} moves electrons in the opposite direction of holes, but Eq.~\eqref{eq: postselbeta} implies that dips and bumps (compared to the NESS) in the effective ``probability field" $\bs{\beta}$ \emph{both} move to the left.
As a density matrix, we then have
\begin{equation}
    \ket{\psi}\bra{\psi} = \sum_{j,k=1}^N b_{jk}(\tau)c_j\rho_N\cdag_k
\end{equation}
where we read off $b_{jk}(\tau) := \beta_j(\tau)\beta^*_k(\tau)$ so that $b_{jj}(\tau) = |\beta_j(\tau)|^2$. Comparing with Eq.~\eqref{eq:time_evolution_of_perturbation}, we therefore see that an initial density bump at position $i$, corresponding to $b_{ii} < b^\mathrm{NESS}_{ii}$, moves to the left.

The above discussion provides the shortest-route argument to highlight the existence of the various effects we predict. One might however prefer to see this straight from the open quantum system description, without referring to the effective Hamiltonian. This is the subject of the next section.
\subsection{Full open quantum system treatment}
Here we derive the forms of the Hamiltonian and jump operators required for the third quantization formalism~\cite{Prosen_2008} on our simple model. Majorana operators are defined via 
\begin{equation}
    c_i = \frac{1}{2}(\gamma_{i,1} - \mathrm{i}\gamma_{i,2}), \quad \{\gamma_{i,\mu}, \gamma_{j,\nu}\} = 2\delta_{ij}\delta_{\mu\nu},
\end{equation}
which generate a Clifford algebra.
We begin with the Hermitian part of the Hamiltonian in~\eqref{eq:hatano_nelson}, which is 
\begin{equation}\label{eq:hermitean_hamiltonian_part}
    H = t\sum_i (\cdag_{i+1}c_i + \cdag_i c_{i+1}) = \frac{t}{4}\sum_i\left[ (\gamma_{i+1,1} + \mathrm{i} \gamma_{i+1,2})(\gamma_{i,1} - \mathrm{i} \gamma_{i,2}) + \text{(h.c.)}\right].
\end{equation}
We wish to use this to obtain the expression of the Majorana Hamiltonian $H$ defined in Eq.~\eqref{eq:majorana_definitions}.
Expanding Eq.~\eqref{eq:hermitean_hamiltonian_part} and using anti-commutation of the Majoranas, half of the terms cancel and we obtain
\begin{equation}
    H = \frac{\mathrm{i}t}{4}\sum_i\left[\gamma_{i+1,2}\gamma_{i,1} - \gamma_{i,1}\gamma_{i+1,2} + \gamma_{i,2}\gamma_{i+1,1} - \gamma_{i+1,1}\gamma_{i,2}\right].
\end{equation}
This specific form is chosen so that the Majorana Hamiltonian is indeed anti-symmetric. In terms of translationally-invariant coefficients $H_{i-j, \mu\nu}$, we may write
\begin{equation}
(H_{1})_{\mu\nu} = \begin{pmatrix}
    0 & -\mathrm{i}t/4 \\
    \mathrm{i}t/4 & 0
\end{pmatrix} = (H_{-1})_{\mu\nu}.
\end{equation}
This allows to easily see how to populate the Hamiltonian matrix as defined in Eq.~\eqref{eq:majorana_definitions} as a function of distance from the main diagonal. For example for $N=2$ fermions, the Majorana Hamiltonian matrix is
\begin{equation}
  H = \begin{bmatrix}
    0 & 0 & 0 & -\im t/4 \\
    0 & 0 & \im t/4 & 0 \\
    0 & -\im t/4 & 0 & 0 \\
    \im t/4 & 0 & 0 & 0
  \end{bmatrix}
\end{equation}
or twice that if we include PBCs; this only occurs at $N=2$, not for larger $N$.
Let us now express the loss operators $L_i = \sqrt{\gamma_l}(c_i - \mathrm{i}c_{i+1})$ as $L_i = \sum_{j,\mu} L_{ij,\mu}\gamma_{j,\mu}$.
In our model, we have
\begin{equation}
    L_i = \frac{\sqrt{\gamma_l}}{2}\left[(\gamma_{i,1} - \mathrm{i} \gamma_{i,2}) - \mathrm{i}(\gamma_{i+1,1} - i\gamma_{i+1,2})\right],
\end{equation}
from which we read off
\begin{equation}\label{eq:jump_matrix}
\begin{aligned}
    L_{i, i, 1} &= \sqrt{\gamma_l}/2, \\
    L_{i, i,2} &= -\mathrm{i}\sqrt{\gamma_l}/2, \\
    L_{i, i+1,1} &= -\mathrm{i}\sqrt{\gamma_l}/2, \\
    L_{i, i+1,2} &= -\sqrt{\gamma_l}/2.
\end{aligned}
\end{equation} 
Analogously, one can obtain such coefficients for the gain operators in Eq.~\eqref{eq:simple_model_jump_operators}.
With these, we may now compute the $2N \times 2N$ matrix $M$ defined in Eq.~\eqref{eq:M_matrix_definition}. Using translationally-invariant notation $(M_a)_{\mu\nu}$ where $a = i-j$, we obtain:
\begin{equation}
    \begin{aligned}
        (M_0)_{\mu\nu} &= \frac{1}{2}\begin{pmatrix}
            \gamma & -\mathrm{i}\delta \\
            \mathrm{i}\delta & \gamma
        \end{pmatrix}, \\
        (M_1)_{\mu\nu} &= \frac{\mathrm{i}}{4}\begin{pmatrix}
            \gamma & -\mathrm{i}\delta \\
            \mathrm{i}\delta & \gamma
        \end{pmatrix}, \\
        (M_{-1})_{\mu\nu} &= -\frac{\mathrm{i}}{4}\begin{pmatrix}
            \gamma & -\mathrm{i}\delta \\
            \mathrm{i}\delta & \gamma
        \end{pmatrix}.
    \end{aligned}
\end{equation}
With these, we now have all the input matrices $(H_a)_{\mu\nu}, \; (M_a)_{\mu\nu}$ required to obtain the Fourier transformed matrices $H(k)_{\mu\nu}, \; M(k)_{\mu\nu}$ [see Eq.~\eqref{eq:fourier_transform_definition_lindblad}]. In turn we may compute the $Z(k)_{\mu\nu}$ matrix from Eq.~\eqref{eq:mom_space_Z_matrix_definition}, which gives
\begin{equation}
    Z(k) = \begin{pmatrix}
        -\mathrm{i}\gamma & -2\mathrm{i}t\cos k +\delta\sin k \\
        2\mathrm{i}t\cos k -\delta\sin k & -\mathrm{i}\gamma
    \end{pmatrix}.
\end{equation}
The eigenvalues of this matrix are
\begin{equation}
\begin{aligned}
    \lambda_{k,1} &= -\mathrm{i}\gamma + 2t\cos k + \mathrm{i}\delta\sin k, \\
    \lambda_{k,2} &= -\mathrm{i}\gamma - 2t\cos k -\mathrm{i}\delta\sin k.
\end{aligned}
\end{equation}
These are indeed the eigenvalues of $\mathcal{H}^{\mathrm{eff}}(k)$ and $-\mathcal{H}^{\mathrm{eff}}(-k)^*$, respectively, as shown generally in Eq.~\eqref{eq:Z_eigenvalues_are_H_eff_eigenvalues}. In this model, both of these eigenvalues have the same winding direction. The discussion in the previous section carries over, and we can have all possible cases depending on the imbalance of gain and loss (i.e. the sign of $\delta$).

\subsection{Topological properties of the NESS}
\begin{figure}
    \centering
    \includegraphics[width=0.95\linewidth]{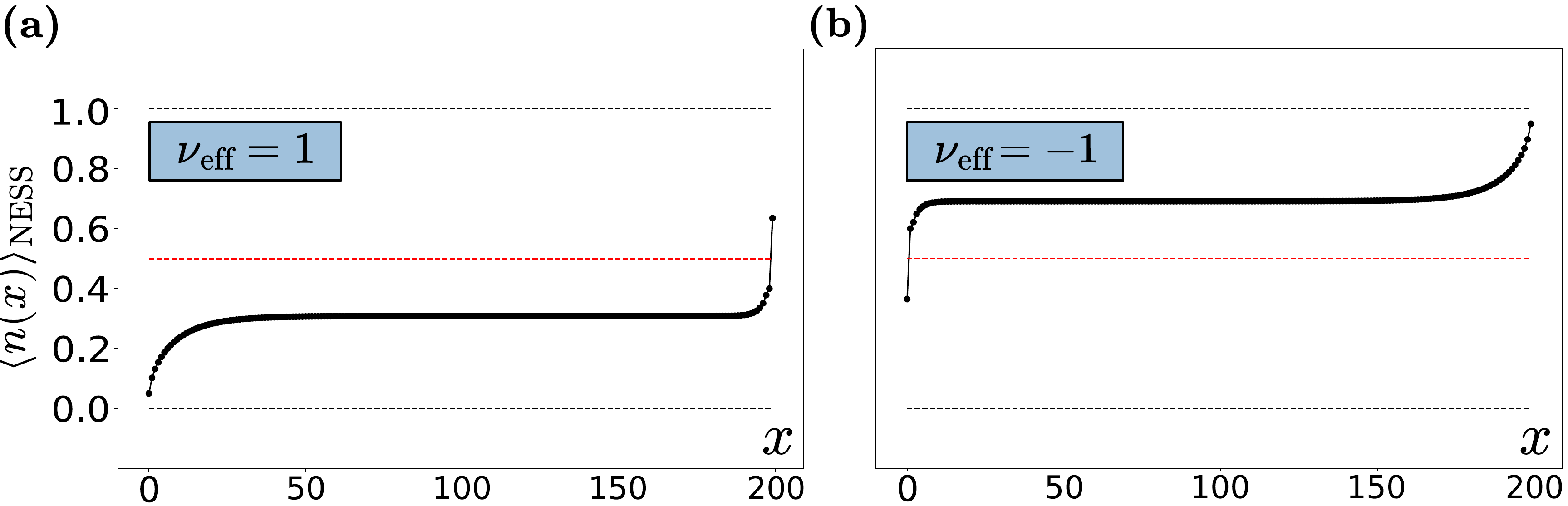}
    \caption{Numerical plots of single-site density expectation values $\langle n(x)\rangle_{\mathrm{NESS}} = \langle \cdag_x c_x \rangle_{\mathrm{NESS}}$ against position $x$ along the chain, in the NESS of the Lindbladian Hatano-Nelson model of Eq.~\eqref{eq:simple_model_jump_operators} for open boundary conditions. (a) NESS density profile for a loss-dominated chain, with parameters $t=5, \gamma_l = 1, \gamma_g = 0.2$. (b) NESS density profile for a gain-dominated chain, with parameters $t=5, \gamma_l=0.2, \gamma_g=1$. In both cases the system size is $N=200$. The winding number sign can be deduced from the bulk occupation being below ($\nu_{\mathrm{eff}}=1$) or above ($\nu_{\mathrm{eff}} = -1$) half-filling (shown as a red horizontal dashed line).}
    \label{fig:NESS_density_profiles}
\end{figure}

We discuss the extent to which the Lindbladian's topology/winding number information can be found in the system's NESS. Fig.~\ref{fig:NESS_density_profiles} shows expectation values of on-site densities along the chain for our loss/gain Hatano-Nelson Lindbladian, both for the loss-dominated/gain-dominated cases. The model's fermionic nature restricts densities to lie between $0$ and $1$. Numerically, we find that whenever the bulk density is below half-filling, the corresponding winding number of the effective Hamiltonian is $\nu_{\mathrm{eff}}=1$, whereas a bulk density above half-filling has $\nu_{\mathrm{eff}}=-1$. Our numerical results imply the simple relation
\begin{equation}
    \nu_{\mathrm{eff}} = \mathrm{sgn}\left(\frac{1}{2}-\ \bar{n}_{\mathrm{NESS}}\right),
\end{equation}
where $\bar{n}_{\mathrm{NESS}}$ is the bulk density in the NESS.

While intuitive, this relation between the NESS and the Lindbladian's topology is model-dependent. Indeed it has been shown~\cite{PhysRevLett.124.040401} that any quadratic Lindbladian can be deformed to have a trivial NESS $\rho_{\mathrm{NESS}}\propto \mathbb1$, without changing the Lindblad spectrum (in particular, without closing the spectral gap) and while preserving all symmetries.

\subsection{Postselection-induced phase transition}
In this section we consider the second model mentioned in the main text, where we change gain operators from $G_i = \sqrt{\gamma_g}(\cdag_i - \mathrm{i}\cdag_{i+1})$ to $G_i = \sqrt{\gamma_g}(\cdag_i + \mathrm{i}\cdag_{i+1})$, with all other data ($H, L_i$) unchanged. This has the effect of flipping the direction of the current caused by the $\sum_i \Gdag_i G_i$ term in $\Hpost$ and $\Heff$. As a result these two Hamiltonians exchange i.e.
\begin{equation}\label{eq:second_gainloss_model_hamiltonians}
\begin{aligned}
  \mathcal{H}^{\mathrm{post}}(k) &= 2t\cos k + \mathrm{i}\delta\sin k -\mathrm{i}\gamma, \\
  \mathcal{H}^{\mathrm{eff}}(k) &= 2t\cos k + \mathrm{i}\gamma\sin k -\mathrm{i}\delta.
\end{aligned}
\end{equation}
When $\gamma_l > \gamma_g$, both $\Hpost$ and $\Heff$ have a skin effect on the right edge (i.e. sites near $i=N$) of the chain, but when $\gamma_g > \gamma_l$, $\Hpost$ has a skin effect on the \emph{left edge}, while $\Heff$ (so also the Lindbladian) retains its skin effect on the right. This means that when postselecting trajectories without quantum jumps (i.e. dynamics under $\Hpost$), for gain dominating loss, the eigenstates localize on the left edge, whereas unconditional dynamics (under $\Heff$) has eigenstates on the right edge. Correspondingly, the point gap in $\Heff$ is always open for any $(\gamma_l, \gamma_g)$, whereas the point gap of $\Hpost$ closes at $\gamma_l = \gamma_g$ and reverses winding direction when going from loss-dominant to gain-dominant régimes, and vice-versa.
\bibliographystyle{unsrt}
\bibliography{refs}